\documentclass[twocolumn]{aastex701}
\usepackage[labelsep=period]{subcaption}

\usepackage{amsmath}
\usepackage{placeins}
\usepackage{array}
\usepackage{tabularx}
\usepackage{CJK}

\begin{document}
\begin{CJK*}{UTF8}{gbsn}

\title{Revisiting Ca\,\textsc{II} Activity Indices in FGK Stars: Systematic Biases in Infrared Triplet Measurements}

\email[]{}
\correspondingauthor{Haotong Zhang}

\author[0009-0006-7506-1299,gname=Xiaozhen,sname=Yang]{Xiaozhen Yang (杨肖振)}
\affiliation{CAS Key Laboratory of Optical Astronomy, National Astronomical Observatories, Chinese Academy of Sciences, Beijing 100101, People's Republic of China}
\affiliation{School of Astronomy and Space Science, University of Chinese Academy of Sciences, Beijing 100049, People's Republic of China}
\email{yangxz@bao.ac.cn}

\author[0000-0002-6506-1985,gname=Xiaoting,sname=Fu]{Xiaoting Fu (符晓婷)}
\affiliation{Purple Mountain Observatory, Chinese Academy of Sciences, Yuanhua road 10, Nanjing 210023, People's Republic of China}
\email{xiaoting.fu@pmo.ac.cn}

\author[0000-0002-5649-7461,gname=Mingjie,sname=Jian]{Mingjie Jian (简明杰)}
\affiliation{Institute of Astronomy, University of Cambridge, Madingley Road, Cambridge CB3 0HA, UK}
\email{jian-mingjie@outlook.com}

\author[0000-0003-2868-8276,gname=Jingkun,sname=Zhao]{Jingkun Zhao (赵景昆)}
\affiliation{National Astronomical Observatories, Chinese Academy of Sciences, Beijing 100101, People's Republic of China}
\email{zjk@nao.cas.cn}

\author[0000-0002-4554-5579,gname=Hailong,sname=Yuan]{Hailong Yuan (袁海龙)}
\affiliation{CAS Key Laboratory of Optical Astronomy, National Astronomical Observatories, Chinese Academy of Sciences, Beijing 100101, People's Republic of China}
\email{yuanhl@bao.ac.cn} 

\author[0000-0003-3884-5693,gname=Zhongrui,sname=Bai]{Zhongrui Bai (白仲瑞)}
\affiliation{CAS Key Laboratory of Optical Astronomy, National Astronomical Observatories, Chinese Academy of Sciences, Beijing 100101, People's Republic of China}
\email{zrbai@nao.cas.cn}

\author[0009-0009-6931-2276,gname=Mengxin,sname=Wang]{Mengxin Wang (汪梦欣)}
\affiliation{CAS Key Laboratory of Optical Astronomy, National Astronomical Observatories, Chinese Academy of Sciences, Beijing 100101, People's Republic of China}
\email{mxwang@nao.cas.cn}

\author[0000-0002-6312-4444,gname=Yiqiao,sname=Dong]{Yiqiao Dong (董义乔)}
\affiliation{CAS Key Laboratory of Optical Astronomy, National Astronomical Observatories, Chinese Academy of Sciences, Beijing 100101, People's Republic of China}
\email{dyq@bao.ac.cn}

\author[0009-0000-6595-2537,gname=Mingkuan,sname=Yang]{Mingkuan Yang (杨明宽)}
\affiliation{CAS Key Laboratory of Optical Astronomy, National Astronomical Observatories, Chinese Academy of Sciences, Beijing 100101, People's Republic of China}
\affiliation{School of Astronomy and Space Science, University of Chinese Academy of Sciences, Beijing 100049, People's Republic of China}
\email{yangmk@bao.ac.cn} 

\author[0009-0009-9065-1846,gname=Ziyue,sname=Jiang]{Ziyue Jiang (蒋子悦)}
\affiliation{CAS Key Laboratory of Optical Astronomy, National Astronomical Observatories, Chinese Academy of Sciences, Beijing 100101, People's Republic of China}
\affiliation{School of Astronomy and Space Science, University of Chinese Academy of Sciences, Beijing 100049, People's Republic of China}
\email{jiangziyue@bao.ac.cn}

\author[0009-0009-0971-5081,gname=Qian,sname=Liu]{Qian Liu (刘倩)}
\affiliation{CAS Key Laboratory of Optical Astronomy, National Astronomical Observatories, Chinese Academy of Sciences, Beijing 100101, People's Republic of China}
\affiliation{School of Astronomy and Space Science, University of Chinese Academy of   Sciences, Beijing 100049, People's Republic of China}
\email{liuqian@bao.ac.cn}

\author[0009-0009-0171-1553,gname=Ganyu,sname=Li]{Ganyu Li (李甘雨)}
\affiliation{CAS Key Laboratory of Optical Astronomy, National Astronomical Observatories, Chinese Academy of Sciences, Beijing 100101, People's Republic of China}
\affiliation{School of Astronomy and Space Science, University of Chinese Academy of Sciences, Beijing 100049, People's Republic of China}
\email{ligy@bao.ac.cn}

\author[0000-0002-6617-5300,gname=Haotong,sname=Zhang]{Haotong Zhang (张昊彤)}
\affiliation{CAS Key Laboratory of Optical Astronomy, National Astronomical Observatories, Chinese Academy of Sciences, Beijing 100101, People's Republic of China}
\email[show]{htzhang@bao.ac.cn}

\begin{abstract}
Synthetic-template subtraction is widely used to measure chromospheric activity in large spectroscopic surveys. However, many solar-like FGK stars show systematically negative Ca\,\textsc{II} infrared triplet (IRT) activity indices. We investigate this effect using solar-like stars from LAMOST DR9, MaStar, and XSL DR3, measuring activity indices ($R^+$) for both the Ca\,\textsc{II} H\&K and IRT lines within a uniform framework. We find that observational effects contribute to the scatter but do not explain the systematic negative bias in $R^+_{\rm IRT}$. This bias mainly arises because radiative-equilibrium photospheric models lack the chromospheric layers, which exhibit a temperature rise and provide additional Ca\,\textsc{II} IRT opacity. This additional opacity makes the observed IRT cores deeper than those in the synthetic photospheric profiles. NLTE effects may contribute but are unlikely to dominate. The different behavior of the H\&K and IRT diagnostics can be related to their different source-function behavior: the stronger chromospheric source-function enhancement of H\&K makes their cores more likely to appear in emission. Increasing the adopted microturbulent velocity partially mitigates the negative offset, serving as an empirical compensation for missing atmospheric physics. In addition, $R^+$ values derived from different synthesis configurations show systematic offsets but generally preserve strong linear correlations, indicating that they can be cross-calibrated. These results clarify the origin of negative Ca\,\textsc{II} IRT residual indices and help interpret template-dependent systematics in chromospheric activity measurements.

\end{abstract}

 \section{Introduction}

In the chromospheres of cool stars, non-radiative heating processes, including magnetic heating, together with reduced radiative cooling, lead to an outward increase in temperature \citep{Linsky2017,Fontenla2016}. 
Under these conditions, chromospheric heating produces emission in the cores of strong spectral lines, such as Ca\,\textsc{II} H\&K, the Ca\,\textsc{II} infrared triplet (IRT), and Mg\,\textsc{II} h and k. 
The strength of this emission reflects the level of chromospheric heating and is therefore widely used as a diagnostic of stellar magnetic activity \citep{Grijs2021}. 

Among these diagnostics, the Ca\,\textsc{II} H\&K and IRT lines are among the most commonly used. 
In vacuum, their central wavelengths are 3934.78~\AA\ (K), 3969.59~\AA\ (H), and 8500.35~\AA, 8544.44~\AA, and 8664.52~\AA\ (IRT). These lines form at different atmospheric heights and exhibit different emission strengths.
Observations and numerical simulations of the quiet Sun indicate that the line cores of the H\&K originate in the upper chromosphere
\citep{Vernazza1981,Bjorgen2018}, 
and the cores of the IRT lines form in the lower chromosphere
\citep{Bjorgen2018,Linsky2017}. 
Enhanced chromospheric heating leads to stronger line-core emission, with the H\&K lines generally showing stronger emission than the IRT \citep{Linsky1979a,Linsky1979b,Strassmeier1990,Pasquini1992}.

Large spectroscopic surveys have widely adopted template-subtraction methods to measure chromospheric activity, because the photospheric contribution can be explicitly removed using synthetic spectra matched to the stellar atmospheric parameters \citep{Andretta2005,Busa2007}. The resulting activity index is defined from the template-subtracted residual core flux and represents the excess chromospheric emission.
This strategy is particularly well suited to survey data, for which empirical calibrations such as the $R'_{\rm HK}$ index \citep{Noyes1984}, derived from the Mount Wilson $S_{\rm MW}$ index \citep{Wilson1968,Duncan1991}, are not always straightforward to apply uniformly across broad ranges of stellar parameters, spectral resolution, and data quality.
The rapid development of synthetic spectral libraries has further enabled such measurements for very large stellar samples. 

However, several survey-based studies have reported negative Ca\,\textsc{II} IRT activity indices in a significant fraction of stars, which is not expected under standard assumptions. For example, \citet{Lanzafame2023} derived IRT activity indices from Gaia DR3 Radial Velocity Spectrometer (RVS) spectra \citep{GaiaMission2016,GaiaDR32023} using synthetic MARCS spectra, while \citet{Huang2024} measured $R^+_{\rm IRT}$ indices from low-resolution spectra in the ninth data release (DR9) of the Large Sky Area Multi-Object Fiber Spectroscopic Telescope (LAMOST; \citealt{2012RAA....12.1197C}) using BT-Settl spectra \citep{Allard2011,Allard2012}. Both studies reported a substantial number of negative values, suggesting that this phenomenon is not specific to a particular survey and that the observed cores can be systematically deeper than those predicted by parameter-matched synthetic spectra. This mismatch appears to be more pronounced in the IRT lines than in H\&K.

The mismatch in the IRT line cores reflects a more general limitation in the correspondence between observed spectra and photospheric templates \citep{Lancon2021}. 
A commonly invoked explanation is that purely radiative-equilibrium photospheric templates lack chromospheric structure, so they can miss the additional absorption that deepens the IRT cores in chromospheric line formation \citep{Cram1979,Lanzafame2023}.
Similar concerns have been noted in Gaia RVS analyses, where the IRT cores were avoided in abundance measurements and activity-related spectral discrepancies were recognized as a source of unreliable parametrization \citep{2023A&A...674A..29R}. Understanding the origin of this core discrepancy is therefore important for interpreting survey-based chromospheric diagnostics from large spectroscopic datasets such as LAMOST and Gaia. More broadly, clarifying why the observed and template spectra diverge in the IRT cores can improve the reliability of template-subtraction-based activity measurements, help explain potential biases in abundance analyses based on these lines, and enable more robust use of the IRT lines in large spectroscopic surveys.

In this work, we investigate the origin of negative Ca\,\textsc{II} IRT indices derived through template-subtraction methods from both the observational and template perspectives and assess how the adopted synthetic templates affect the resulting activity measurements.
Section~\ref{sec:data} introduces the observational samples and synthetic templates. 
Section~\ref{sec:method} describes the activity-index definitions and measurement procedures. 
Section~\ref{sec:lamostidx} presents the LAMOST activity indices. 
Section~\ref{sec:diagnostics} examines the contributions of observational systematics and template-related mismatch. 
Section~\ref{sec:vmic_cor} presents an empirical microturbulence-based mitigation. 
Section~\ref{sec:consistency} investigates the dependence of $R^+$ on the synthesis configurations and the potential for cross-calibration. 
Our conclusions are summarized in Section~\ref{sec:conclusion}.

\section{Observational Data and Synthetic Templates}\label{sec:data}

\begin{figure*}[htbp]
    \centering
    \includegraphics[width=0.95\textwidth]{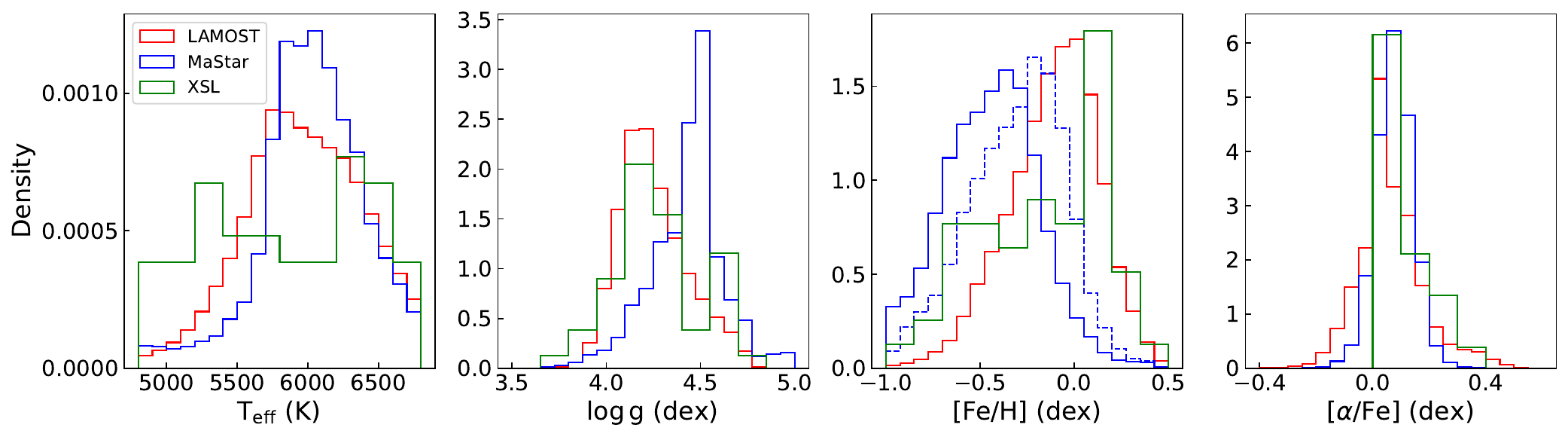}
    \caption{Distributions of $T_{\mathrm{eff}}$, $\log g$, [Fe/H], and [$\alpha$/Fe] for LAMOST DR9, MaStar, and XSL. The blue dashed line marks the median [Fe/H]$_\mathrm{cal}$ of MaStar. Due to the small number of stars in XSL, its bin sizes are twice those of LAMOST and MaStar.}
    \label{fig:3surveypara}
\end{figure*}

This section presents the observational samples used in this work and the synthetic spectra adopted as photospheric templates. 
The three observational data sets serve complementary purposes: LAMOST DR9 provides the primary large-sample data set for statistical analysis, the SDSS DR17 MaNGA Stellar Library (MaStar; \citealt{MaStar,SDSSDR17}) offers an independent survey-level comparison for assessing potential systematics, and the X-shooter Spectral Library (XSL) DR3 \citep{XSLDR3,Xshooter2011} provides higher-resolution spectra for examining observed--template mismatches in line profiles. 
An overview of the observational data sets is given in Table~\ref{tab:dataset_summary}, and Figure~\ref{fig:3surveypara} shows the atmospheric-parameter distributions of the selected stars. 
These overview summaries provide context for the more detailed descriptions in the following subsections.

\begin{table*}[htbp]
\centering
\caption{Summary of the observational data sets used in this work}
\label{tab:dataset_summary}
\begin{tabular}{lccccc}
\hline
\hline
Survey/Data Release & Spectral Coverage & Resolving Power & Wavelength Sampling  & SNR Criterion & $N$ \\
  & (HK / IRT) & ($\lambda/\Delta\lambda$) & (\AA) &  &  \\
\hline
LAMOST DR9 LRS
& HK + IRT
& $\sim$1310 (HK), $\sim$1910 (IRT)
& $\sim$1 (HK), $\sim$2 (IRT)
& SNR$_r \geq 50$
& 428\,253 \\

MaStar SDSS DR17
& HK + IRT
& $\sim$1090 (HK), $\sim$2010 (IRT)
& $\sim$1 (HK), $\sim$2 (IRT)
& $\langle{\rm SNR}\rangle \geq 100$
& 7\,336 \\

XSL DR3
& HK + IRT
& 9774
& 0.1
& ---
& 52 \\
\hline
\end{tabular}
\end{table*}

\subsection{LAMOST DR9 Low-Resolution Spectra}\label{subsec:lamost_data}

We selected solar-like stars, including F-, G-, and K-type stars, from the LAMOST DR9 v2.0 low-resolution spectroscopic survey (LRS). 
The sample selection follows \citet{Huang2024}, based on the parameter-space criteria of \citet{Zhang2022} (see their Figure~3), which use the stellar atmospheric parameters ($T_{\mathrm{eff}}$, $\log g$, and [Fe/H]) together with the spectral signal-to-noise ratio (SNR).
To ensure reliable activity measurements in both the blue (H\&K) and near-infrared (IRT) spectral regions, we further impose band-dependent SNR constraints. 
According to the LAMOST LRS Stellar Parameter Catalog of A, F, G and K Stars (AFGK catalog)\footnote{\url{https://www.lamost.org/dr9/v2.0/catalogue}}, the uncertainties in the stellar atmospheric parameters decrease significantly for spectra with $\mathrm{SNR}_r \geq 50$ (signal-to-noise ratio in the $r$ band).
We therefore adopt $\mathrm{SNR}_r \geq 50$ as the baseline quality criterion.

We further restrict the metallicity range to $-1 \leq \mathrm{[Fe/H]} \leq 0.5$. No explicit constraint is imposed on [$\alpha$/Fe], which is used here as a proxy for the overall $\alpha$-element enrichment relevant to the formation of the Ca\,\textsc{II} lines. When [$\alpha$/Fe] is unavailable in the AFGK catalog, we estimate it empirically from [Fe/H] as
\begin{equation}
[\alpha/\mathrm{Fe}] =
\begin{cases}
0, & \mathrm{[Fe/H]} \ge 0 \\[4pt]
-0.4 \times \mathrm{[Fe/H]}, & -1 \le \mathrm{[Fe/H]} < 0 \; .
\end{cases}
\end{equation}
This empirical prescription is intended to approximate the mean trend of [$\alpha$/Fe] with [Fe/H] in Galactic stellar populations \citep{GESAlphaFe,APOGEE2015alphafe,Khoperskov2021}.
Using the unique LAMOST source identifiers provided in the AFGK catalog, we obtained spectra of 428\,253 solar-like stars. Figure~\ref{fig:3surveypara} shows the distributions of $T_{\mathrm{eff}}$, $\log g$, [Fe/H], and [$\alpha$/Fe] for these stars.

\begin{figure}[htbp]
    \centering
    \includegraphics[width=0.95\columnwidth]{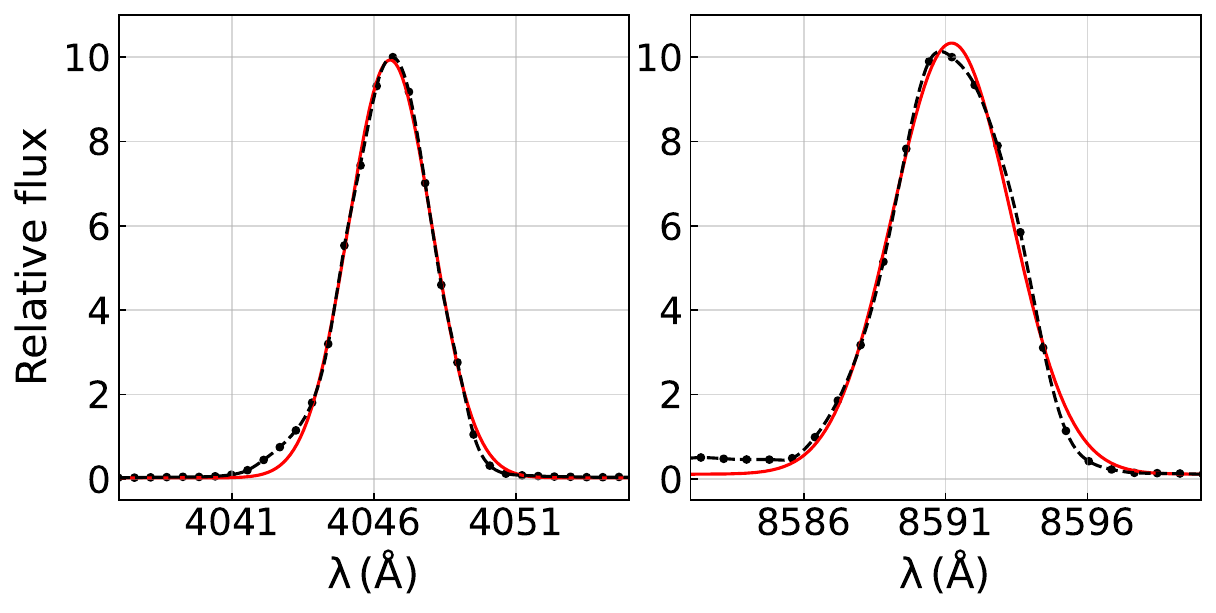}
    \caption{LAMOST arc-lamp emission lines. The left panel shows the profile of the 4046.6~\AA\ line and the right panel shows the profile of the 8591.3~\AA\ line. Black points: arc-lamp spectrum; black dashed curve: cubic spline interpolation; red curve: fitted Gaussian.}
    \label{fig:LSFprofile}
\end{figure}

The LAMOST DR9 low-resolution spectra have a nominal resolving power of $\lambda/\Delta\lambda \approx 1800$ at 5500~\AA\ \citep{Lamost2012}, consistent with the value adopted by \citet{Huang2024}. Although the instrumental $\Delta\lambda$ is expected to be approximately constant, we do not assume a single constant value. Instead, we characterize the wavelength-dependent line-spread function (LSF) and estimate $\Delta\lambda$ using LAMOST arc-lamp spectra (\citealt{Liu_2026}). Deviations from a purely Gaussian shape are present in some arc-lamp emission lines, as illustrated in Figure~\ref{fig:LSFprofile}; we therefore represent the LSF using the emission-line profile after cubic-spline interpolation, while $\Delta\lambda$ is characterized by a Gaussian fit. Because the central question concerns subtle line-core mismatches, the convolution kernel itself must be validated rather than assumed Gaussian. Unless otherwise specified, the synthetic templates are convolved with the instrumental LSF to match the resolution of the LAMOST spectra.

\subsection{MaStar and XSL Spectra}\label{subsec:mastar_xsl}
In addition to LAMOST DR9, we include solar-like stars from SDSS DR17 MaStar and XSL DR3 as independent reference samples to analyze potential biases arising from reliance on a single survey.

For MaStar, we select the subset corresponding to the 95.5th-percentile resolution curve, which contains the largest number of stars, and further require a mean signal-to-noise ratio of $\langle\mathrm{SNR}\rangle \geq 100$. 
This subset has low spectral resolution, as summarized in Table~\ref{tab:dataset_summary}, comparable to that of the LAMOST low-resolution spectra, making it well suited for an independent survey-level comparison.
MaStar also provides stellar atmospheric parameters derived using four independent methods, together with their median values \citep{Imig2022,Lazarz2022,Hill2022_teffloggfeh}.
In this work, we adopt the median values of $T_{\mathrm{eff}}$, $\log g$, [Fe/H], and [$\alpha$/Fe], as well as the median calibrated metallicity, [Fe/H]$_{\mathrm{cal}}$, obtained through cross-matching with APOGEE \citep{APOGEE_Overview,APOGEE_ASPCAP}. 
A total of 7336 solar-like stars were selected from MaStar.

The XSL DR3 spectra have a resolving power of $\lambda/\Delta\lambda = 9774$ and have been corrected for Galactic extinction. 
Their substantially higher spectral resolution enables more informative line-profile comparisons than the survey spectra alone. $T_{\mathrm{eff}}$, $\log g$, and [Fe/H] are adopted from \citet{Arentsen2019}, while [$\alpha$/Fe] values are taken from \citet{Santos-Peral2023} and Gaia DR3. 
Given the limited sample size and the generally high signal-to-noise ratios of the XSL spectra, no additional SNR requirement is imposed. 
In total, 52 solar-like stars were included from XSL.

\subsection{Synthetic Spectra as Photospheric Templates}\label{sec:temspec}

We use both synthetic spectra computed in this work and publicly available synthetic spectral libraries to construct the photospheric templates. All spectra are computed using one-dimensional stellar atmospheres under the local thermodynamic equilibrium (LTE) assumption and have very high intrinsic spectral resolution  ($\lambda/\Delta\lambda \geq 300{,}000$). Together, these templates span broad grids in $T_{\mathrm{eff}}$, $\log g$, [Fe/H], and [$\alpha$/Fe].

\begin{table}[htbp]
\centering
\caption{Mapping of Synthesis Configurations}
\label{tab:config_mapping_separated}
\setlength{\tabcolsep}{6pt}
\renewcommand{\arraystretch}{1.05}

\begin{tabular}{lp{0.5\columnwidth}}
\hline
\hline
Label & Code + Model Atmosphere \\
\hline
A & SPECTRUM + MARCS \\
B & SPECTRUM + ATLAS9 \\
C & Turbospectrum + MARCS \\
D & Turbospectrum + ATLAS9 \\
E & MOOG + MARCS \\
F & MOOG + ATLAS9 \\
G & SYNTHE + MARCS \\
H & SYNTHE + ATLAS9 \\
I & SME + MARCS \\
J & SME + ATLAS9 \\
\hline
\end{tabular}

\vspace{0.5em}
\begin{tabular}{lp{0.5\columnwidth}}
\hline
\hline
Label & Abundance Scale + Line List \\
\hline
1 & Grevesse2007 + VALD \\
2 & Grevesse2007 + GESv6 \\
3 & Asplund2009 + VALD \\
4 & Asplund2009 + GESv6 \\
\hline
\end{tabular}

\tablecomments{Radiative transfer codes: SPECTRUM \citep{SPECTRUM}, MOOG \citep{MOOG}, Turbospectrum \citep{Turbospectrumpaper,Turbospectrumcode}, SYNTHE \citep{SYNTHEcode}, and SME \citep{SMEcode}. Model atmospheres: MARCS \citep{Gustafsson2008} and ATLAS9 \citep{Castelli2003}. Solar abundance scales: \citet{Grevesse2007} and \citet{Asplund2009}. Line lists: VALD \citep{VALD2011,2015PhyS...90e4005R} and GESv6 \citep{GESlinelist}.}
\end{table}

The synthetic spectra computed in this work are generated with \texttt{iSpec} \citep{iSpec2014,iSpec2019}, which provides access to multiple radiative transfer codes, model atmospheres, solar abundance scales, and atomic and molecular line lists. 
In particular, we adopt the solar abundance scales of \citet{Grevesse2007} and \citet{Asplund2009} (hereafter Grevesse2007 and Asplund2009). 
The stars analyzed in this work are dwarfs with high surface gravities, $\log g \simeq$ 3.5--5.0; for this high-gravity range, the MARCS and ATLAS9 model atmospheres used in our \texttt{iSpec} syntheses are plane-parallel.
By combining these components, we construct multiple spectral sets spanning a range of synthesis configurations. As summarized in Table~\ref{tab:config_mapping_separated}, each configuration is identified by a letter--number label, where the letter denotes the combination of radiative transfer code and model atmosphere, and the number specifies the adopted solar abundance scale and line list.

Unless otherwise noted, we adopt the A1 template set as the fiducial template set throughout this work. 
This configuration uses the SPECTRUM radiative transfer code, MARCS model atmospheres, the Grevesse2007 solar abundances, and the VALD line list, which covers the wavelength range 300--1100~nm. 
In this setup, the Ca abundance follows the adopted [$\alpha$/Fe] scaling. 
The microturbulent and macroturbulent velocities ($V_{\mathrm{mic}}$ and $V_{\mathrm{mac}}$) are estimated as functions of $T_{\mathrm{eff}}$, $\log g$, and [Fe/H] following \citet{Gaiabenchmark2014} and are used in the spectral synthesis. 

We also use three publicly available synthetic spectral libraries: BT-Settl \citep{Allard2011,Allard2012}, the G\"ottingen Spectral Library (GSL; \citealt{Husser2013}), and the NewEra library \citep{Hauschildt2025}. 
\added{These libraries were all computed with the PHOENIX stellar-atmosphere code \citep{PHOENIXcode}, assuming spherical symmetry and adopting the Asplund2009 solar abundance scale.}
For stars with high surface gravities, the difference between spherical and plane-parallel geometries is expected to be small, since sphericity effects mainly become important for low-gravity models \citep{Gustafsson2008}. Moreover, the PHOENIX-based libraries are used here primarily as comparison template sets, whereas the fiducial results are based on the A1 templates; this geometrical difference therefore does not affect the main conclusions of this work.
The GSL was generated with PHOENIX version~16 using the ACES equation of state \citep{Barman2011}. 
The NewEra was computed with PHOENIX version~20, also using ACES, but with substantial updates to the atomic and molecular databases relative to GSL.

\section{Methods}\label{sec:method}

This section defines the activity indices used in this work and summarizes the measurement procedure. 
We first define the Ca~\textsc{II} activity indices, then describe the processing steps used to construct parameter-matched photospheric templates and measure the residual core flux, and finally present the local normalization procedure.

\subsection{Activity Indices}\label{subsec:actindex}
Following \citet{Lanzafame2023} and \citet{Huang2024}, we measure the Ca~\textsc{II} IRT activity indices and extend the same definition to the H\&K lines.
For each Ca~\textsc{II} line, we define the line-core index as the mean normalized flux over the line-core window \citep{2011arXiv1107.5325L}:
\begin{equation}
R = \frac{1}{2\Delta\lambda}\int_{\lambda_0-\Delta\lambda}^{\lambda_0+\Delta\lambda}
\frac{F(\lambda)}{C(\lambda)}\, \mathrm{d}\lambda,
\end{equation}
where $\lambda_0$ is the line center, $\Delta\lambda = 0.5$~\AA, $F(\lambda)$ is the flux spectrum, and $C(\lambda)$ is the fitted local pseudo-continuum.
The normalized spectrum is therefore given by $F(\lambda)/C(\lambda)$.
For an equivalent width measured over the same fixed core window, the two quantities are related by
\begin{equation*}
R = 1 - \frac{W_{\rm core}}{2\Delta\lambda}.
\end{equation*}
A larger absorption equivalent width corresponds to a smaller value of $R$.

The template-subtracted residual core-flux index is defined as
\begin{equation}
\begin{aligned}
R^+ &= \frac{1}{2\Delta\lambda}\int_{\lambda_0-\Delta\lambda}^{\lambda_0+\Delta\lambda}
\left[
\frac{F_{\rm obs}(\lambda)}{C_{\rm obs}(\lambda)} -
\frac{F_{\rm tem}(\lambda)}{C_{\rm tem}(\lambda)}
\right] \mathrm{d}\lambda \\
&= R_{\rm obs} - R_{\rm tem},
\end{aligned}
\label{eq:Rplus}
\end{equation}
where the subscripts ``obs'' and ``tem'' denote the observed and parameter-matched template spectra, respectively.
Therefore, $R^+$ is closely related to an excess equivalent width, but is expressed as a dimensionless mean residual core flux over the fixed window. With this sign convention, positive $R^+$ indicates excess core flux relative to the template, corresponding to a shallower absorption core or stronger emission filling.

Following the definition of the Mount Wilson $S_{\rm MW}$ index, $R^+_{\rm HK}$ is defined as the average of the Ca~\textsc{II} H and K indices.
For the IRT, we measure indices for the three lines at $\lambda8498$, $\lambda8542$, and $\lambda8662$, which are collectively denoted as $R^+_{\rm IRT}$. Among them, $R^+_{8542}$ is commonly used as a representative IRT activity index \citep{Mittag2013,Huang2024}.

\subsection{Measurement Procedure}\label{sec:dataprocess}

The activity indices are measured using the following procedure. 
First, all spectra are converted to vacuum wavelengths, adopting the air-to-vacuum conversion of \citet{Ciddor1996}, and radial velocity (RV) corrections are applied to the observed spectra. 
Second, synthetic spectra are interpolated over a grid in $T_{\mathrm{eff}}$, $\log g$, [Fe/H], and [$\alpha$/Fe] to construct parameter-matched photospheric templates. 
Third, the high-resolution synthetic templates are convolved to match the instrumental resolution of the observed spectra. 
Finally, both the observed and template spectra are locally normalized (see Section~\ref{sec:normalization}), and the activity indices are computed by integrating the normalized line profiles using cubic-spline interpolation.

For the low-resolution LAMOST and MaStar spectra, the wavelength sampling interval in the IRT region is approximately 2~\AA\ (see Table~\ref{tab:dataset_summary}). 
Because our adopted index uses $\Delta\lambda=0.5$~\AA, the corresponding total integration width is 1~\AA. 
\citet{Huang2024} compared measurements obtained with total integration widths of 1 and 2~\AA\ and found the resulting differences to be negligible. 
We independently verified this point and found that, although the absolute values of $R^+$ change slightly with the adopted width, the qualitative behavior and the conclusions of this work remain unchanged.

\begin{figure*}[htbp]
    \centering
    \begin{subfigure}{0.98\textwidth}
        \includegraphics[width=\textwidth]{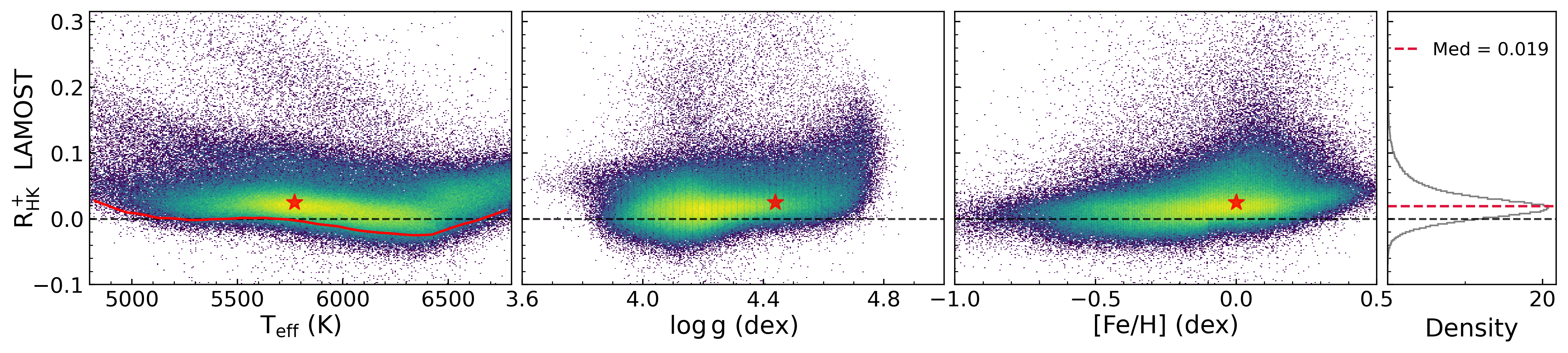}
        \vspace{-0.55cm}
    \end{subfigure}
    \begin{subfigure}{0.98\textwidth}
        \includegraphics[width=\textwidth]{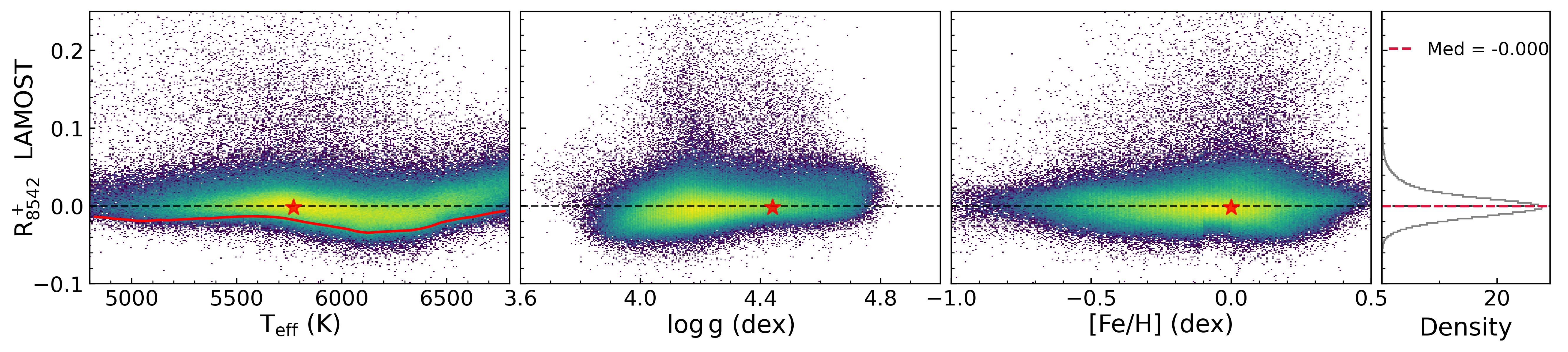}
    \end{subfigure}
    \caption{$R^+_{\rm HK}$ and $R^+_{8542}$ as functions of $T_{\rm eff}$, $\log g$, and [Fe/H] for the LAMOST sample, derived using the A1 templates with $V_{\rm mic} = V^{\rm t}_{\rm mic}$. The upper panels show $R^+_{\rm HK}$, and the lower panels show $R^+_{8542}$. The red star marks the Sun, for which $R^+_{\rm HK}=0.0252$ and $R^+_{8542}=-0.0018$. In the $T_{\rm eff}$ panels, the red curve denotes the 5th-percentile relation. High-density regions are shown in yellow, and the median lines are plotted in red. The black horizontal dashed line indicates $R^+ = 0$.}
    \label{fig:Rpluslamost}
\end{figure*}

\subsection{Local Normalization}\label{sec:normalization}

Accurate local normalization is particularly important in our work. 
To minimize differential biases, we apply the same local normalization procedure to both the observed and template spectra.
We use a two-step pseudo-continuum fitting scheme. 
First, a 250~\AA\ window centered on the Ca~\textsc{II} H\&K or IRT region is selected, and prominent absorption features, including the diagnostic lines themselves, are masked. 
A fourth-order polynomial is then fitted to the remaining points to provide an initial estimate of the local pseudo-continuum. 
The spectra are normalized by this initial fit, and the normalized flux values are sorted in descending order. 
The upper one-third of the points, corresponding to flux levels closest to unity, are retained as candidate continuum points.  
A second fourth-order polynomial is then fitted to the selected continuum points, yielding the final locally normalized spectra used for the index measurements.

\section{Negative $R^+$ Values in LAMOST}\label{sec:lamostidx}

We first measured the $R^+_{\rm HK}$ and $R^+_{8542}$ indices for the LAMOST sample using the A1 templates in order to characterize the occurrence of negative values. For each star, the template spectrum was interpolated to the corresponding LAMOST atmospheric parameters and then degraded to the instrumental resolution using LSF convolution rather than Gaussian convolution, after which the activity indices were calculated. The LAMOST results are presented in Figure~\ref{fig:Rpluslamost}.

In our measurements, the median value of $R^+_{8542}$ is 0, and more than half of the stars have $R^+_{8542} < 0$, indicating a substantial population of negative values, consistent with the results of \citet{Lanzafame2023} and \citet{Huang2024}. These negative values imply that the observed line cores are deeper than those of the corresponding templates. By contrast, $R^+_{\rm HK}$ was not evaluated in those studies. In our sample, most stars show positive $R^+_{\rm HK}$ values, with a higher median value of 0.019, consistent with chromospheric heating in FGK stars. For both $R^+_{\rm HK}$ and $R^+_{8542}$, more negative values are preferentially found among stars with higher effective temperatures and lower surface gravities. 
If the 5th percentile is adopted as an empirical lower envelope, denoted by $R^+_{\rm lower}$, it provides a robust proxy that is less sensitive to outliers and measurement scatter.
We therefore next systematically examine the observational and template-related origins of the negative $R^+$ values.


\section{Diagnostics of Negative $R^+$ Values} \label{sec:diagnostics}

In our framework, $R^+$ quantifies the difference between the observed line-core flux and that predicted by a photospheric template. If both the template physics and measurement procedure were perfect, inactive stars would cluster near $R^+ \approx 0$, while chromospheric activity would yield positive values. In practice, however, $R^+$ is a template-dependent statistic; it is sensitive to observational systematics and, crucially, to the physical assumptions inherent in the synthetic templates. Consequently, systematically negative residual indices ($R^+$) do not imply unphysical negative emission, but rather indicate that the adopted template overpredicts the line-core flux. In this section, we treat the prevalence of negative $R^+$ as a diagnostic of potential biases. We assess observational and template-related contributions in turn to determine which effects account for the observed offsets.

\begin{figure*}[htbp]
    \centering
    \includegraphics[width=0.98\textwidth]{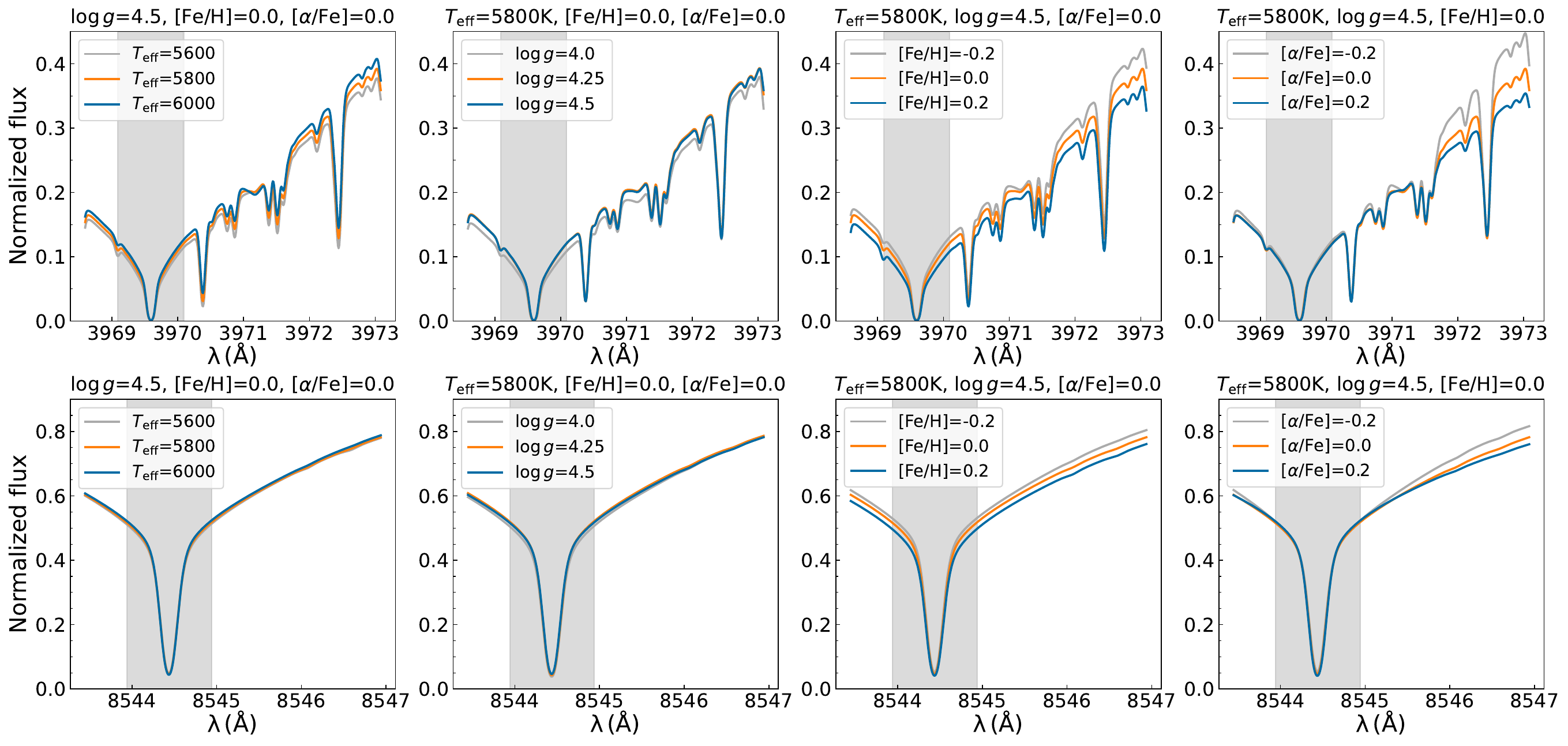}
   \caption{Variations in the normalized Ca\,\textsc{II} H\&K and IRT line profiles as a function of $T_{\rm eff}$, $\log g$, [Fe/H], and [$\alpha$/Fe], using synthetic spectra computed with the A1 configuration. The upper panels show the Ca\,\textsc{II} H line, and the lower panels show the $\lambda8542$ line. Fixed parameters are listed in the panel titles, while the varying parameter in each panel is indicated by color. The adopted parameter steps are $\Delta T_{\rm eff} = 200$~K, $\Delta \log g = 0.25$~dex, $\Delta$[Fe/H]$=0.2$~dex, and $\Delta[\alpha/\mathrm{Fe}] = 0.2$~dex. $V_{\mathrm{mic}}$ and $V_{\mathrm{mac}}$ are held fixed. The gray shaded region marks the 1~\AA\ line-core window. All wavelengths are given in vacuum, and the spectral resolving power is 300,000.}
    \label{fig:para_influ_profile}
\end{figure*}

\subsection{Assessment of Observational Systematics}\label{subsec:obsissue}

To assess whether observational effects contribute to the negative $R^+$ values, we analyze spectra from LAMOST DR9, MaStar, and XSL using a uniform analysis procedure. 
We examine whether the effect can be explained by survey-dependent offsets in the stellar atmospheric parameters (Section~\ref{subsubsec:para}), by the treatment of the instrumental LSF (Section~\ref{subsubsec:LSF}), or by the combined measurement uncertainties (Section~\ref{subsubsec:Error}). For comparison with LAMOST, we also present the $R^+$ results for MaStar and XSL later in this subsection.

\subsubsection{Atmospheric Parameters}\label{subsubsec:para}

Accurate stellar atmospheric parameters are essential for template-based activity measurements, because the photospheric template is constructed from the adopted $T_{\mathrm{eff}}$, $\log g$, [Fe/H], and [$\alpha$/Fe]. Systematic offsets or random uncertainties in these parameters can therefore introduce template mismatch, directly bias $R_{\mathrm{tem}}$, and propagate into the derived $R^{+}$ indices.

We first examine the effects of $T_{\mathrm{eff}}$, $\log g$, [Fe/H], and [$\alpha$/Fe] on the line profiles using \texttt{iSpec}, adopting synthesis settings identical to those of the A1 configuration. Figure~\ref{fig:para_influ_profile} shows the normalized templates for different stellar parameters. The H\&K and IRT profiles exhibit different sensitivities in their wings and cores. Both the line cores and wings are strongly sensitive to [Fe/H], while [$\alpha$/Fe] mainly affects the wings; in both cases, the normalized flux decreases as [Fe/H] and [$\alpha$/Fe] increase. By contrast, changes in $T_{\mathrm{eff}}$ and $\log g$ produce only minor variations in the profiles. This dominant dependence on abundance parameters implies that uncertainties or systematic offsets in [Fe/H] and [$\alpha$/Fe] can significantly bias the derived indices.

\begin{figure*}[htbp]
    \centering
    \includegraphics[width=0.98\textwidth]{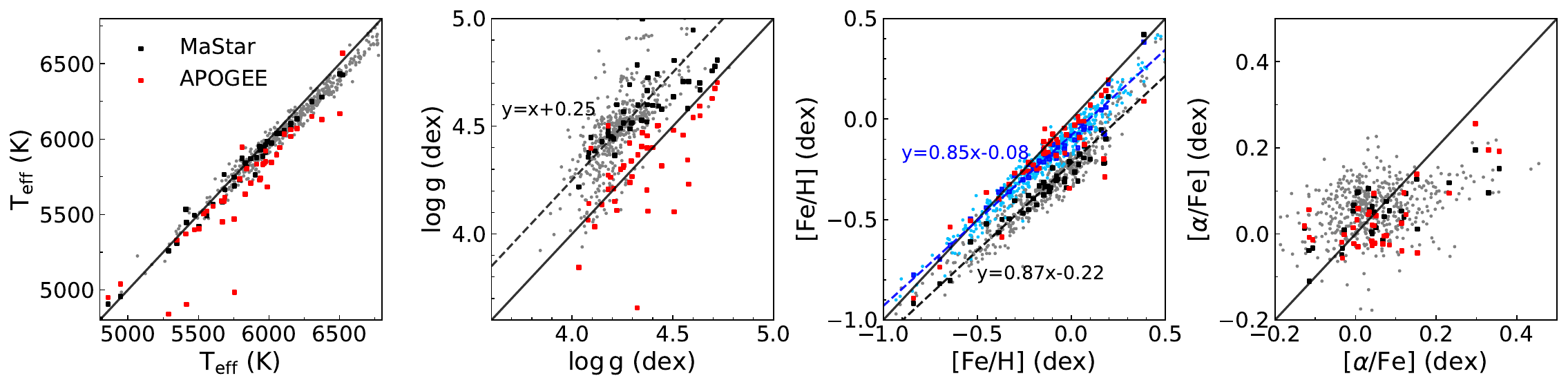}
    \caption{Atmospheric-parameter comparison for cross-matched samples. LAMOST parameters are shown on the x-axis, while the corresponding MaStar and APOGEE parameters are shown on the y-axis. Gray: LAMOST--MaStar cross-match. Large squares: LAMOST--MaStar--APOGEE triple match, with APOGEE parameters in red and MaStar median parameters in black. Blue points in [Fe/H] show MaStar [Fe/H]$_\mathrm{cal}$. Solid black line: $y=x$; dashed line: estimated relation between MaStar and LAMOST parameters.}
    \label{fig:3surveypara_crossmatch}
\end{figure*}

\begin{figure*}[htbp]
    \begin{subfigure}[t]{0.45\textwidth}
        \centering
        \includegraphics[width=\textwidth]{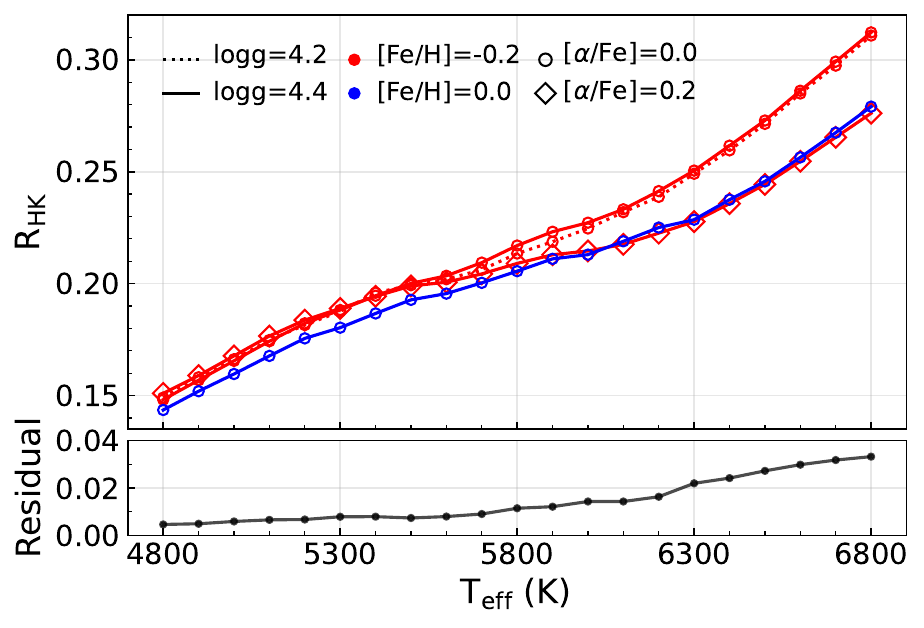}
    \end{subfigure}
    \hfill
    \begin{subfigure}[t]{0.45\textwidth}
        \centering
        \includegraphics[width=\textwidth]{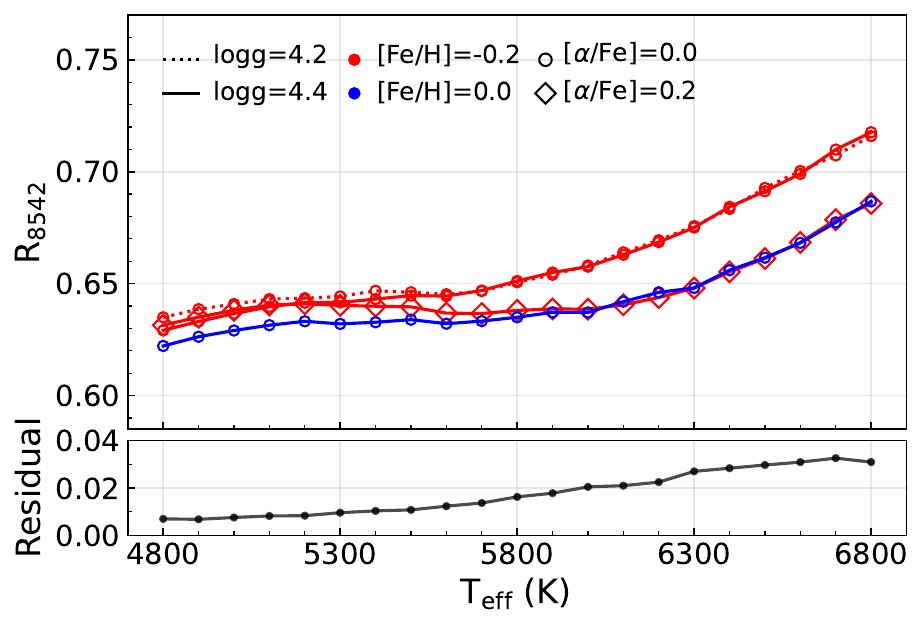}
    \end{subfigure}
    \caption{Variations of $R_{\rm HK}$ (left panel) and $R_{8542}$ (right panel) with $T_{\rm eff}$, $\log g$, [Fe/H], and [$\alpha$/Fe] for the A1 templates at LAMOST resolution. Different line styles denote different $\log g$ values, colors denote different [Fe/H] values, and marker types denote different [$\alpha$/Fe] values. The residuals show $\Delta R$ between two [Fe/H] values at $\log g = 4.4$ and [$\alpha$/Fe]$=0$.}
    \label{fig:para_influ_R}
\end{figure*}

Figure~\ref{fig:3surveypara_crossmatch} shows the LAMOST--MaStar cross-matched sample and reveals systematic differences in both $\log g$ and [Fe/H] between the two surveys. The [Fe/H] values from LAMOST are approximately 0.2~dex higher than the MaStar median [Fe/H] values and about 0.06~dex higher than [Fe/H]$_{\mathrm{cal}}$. For stars with $\log g > 4.2$, the LAMOST $\log g$ values are lower than the MaStar values by approximately 0.25~dex. Because [Fe/H] affects the line profiles more strongly than $\log g$, systematic metallicity offsets are expected to propagate more directly into $R_{\mathrm{tem}}$ and hence into $R^{+}$.

Figure~\ref{fig:para_influ_R} illustrates how the A1-template $R_{\rm HK}$ and $R_{8542}$ indices depend on stellar atmospheric parameters at LAMOST resolution. A decrease of 0.2~dex in either [Fe/H] or [$\alpha$/Fe] can increase both $R_{\rm HK}$ and $R_{8542}$, reaching as much as 0.03 at high $T_{\mathrm{eff}}$. By contrast, changes of 100~K in $T_{\mathrm{eff}}$ or 0.2~dex in $\log g$ have only a minor effect, with a maximum variation of about 0.013 even at high $T_{\mathrm{eff}}$. If [Fe/H] is underestimated by 0.2~dex, the matched template yields a larger $R_{\mathrm{tem}}$ and therefore reduces $R^{+}$ by as much as 0.03, according to Equation~(\ref{eq:Rplus}). In principle, an underestimation of [Fe/H] or [$\alpha$/Fe] by $\sim$0.2~dex could account for part of the distribution extending to negative $R^{+}$ values in the LAMOST sample, although such large systematic offsets are unlikely.

By comparison with FGK stellar parameters in the PASTEL catalog \citep{Soubiran2016} and with cluster members, \citet{Soubiran2022} evaluated the offsets and precisions of [Fe/H] measurements in several spectroscopic surveys, including LAMOST DR5 \citep{LamostDR5data} and APOGEE DR16 \citep{APOGEEDR16}. For metal-poor stars, these surveys tend to overestimate [Fe/H] relative to PASTEL by 0.06--0.18~dex, with APOGEE DR16 showing the smallest offset. APOGEE also exhibits low dispersion for cluster members, indicating excellent performance. By cross-matching LAMOST DR9, MaStar, and APOGEE DR16, we obtained a sample of 43 stars. As shown in Figure~\ref{fig:3surveypara_crossmatch}, the LAMOST parameters are more consistent with those from APOGEE, whereas the MaStar parameters, particularly the median [Fe/H], require further correction. We therefore conclude that atmospheric-parameter biases are unlikely to be the primary cause of the systematically negative $R^+_{8542}$ values in the LAMOST sample.

\begin{figure*}[htbp]
    \centering
    \begin{subfigure}{0.98\textwidth}
        \includegraphics[width=\textwidth]{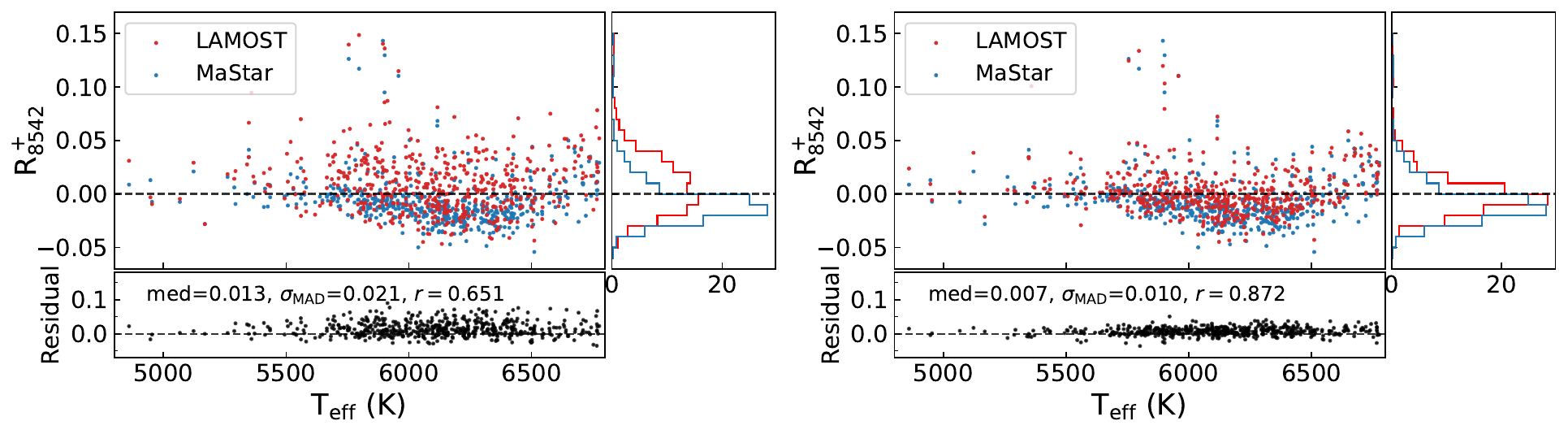}
    \end{subfigure}
    \caption{Comparison of $R^+_{8542}$ measurements for the cross-matched LAMOST and MaStar samples under two convolution schemes. The left panel shows the case in which Gaussian convolution is applied to both datasets, while the right panel shows the case in which the LAMOST spectra are convolved with the instrumental LSF and the MaStar spectra with a Gaussian kernel. In each panel, the upper subpanel presents the measured $R^+_{8542}$ values as a function of $T_{\mathrm{eff}}$, and the lower subpanel shows the residuals, defined as $R^+_{8542,\mathrm{LAMOST}}-R^+_{8542,\mathrm{MaStar}}$. The median, scaled median absolute deviation ($1.4826\times\mathrm{MAD}$), and Pearson correlation coefficient $r$ are labeled in the lower subpanels. The gray dashed line marks the zero level.}
    \label{fig:sample_useLSF8542}
\end{figure*}

\subsubsection{LSF and Gaussian Convolution}\label{subsubsec:LSF}

Gaussian convolution is commonly used to degrade high-resolution templates to the resolution of observed spectra. However, for realistic forward modeling and activity-index measurements, convolution with the instrumental LSF is more appropriate. As shown in Figure~\ref{fig:LSFprofile}, the LAMOST arc-lamp emission line is not always well described by a Gaussian profile and can exhibit noticeable departures from Gaussianity, including a lower peak and non-Gaussian wings.

To assess the impact of the adopted convolution kernel, we used the LAMOST--MaStar cross-matched sample and adopted the stellar parameters from the LAMOST AFGK catalog. For each LAMOST spectrum, the A1 template was convolved either with a Gaussian kernel or with the instrumental LSF corresponding to the nearest observing epoch. For the MaStar spectra, Gaussian convolution was applied in both cases. Figure~\ref{fig:sample_useLSF8542} presents the comparison for $R^+_{8542}$; the corresponding changes in $R^+_{\mathrm{HK}}$ are described quantitatively below.

When Gaussian convolution is applied to both surveys, the residuals between the LAMOST and MaStar measurements show a systematic positive offset. After replacing the Gaussian kernel with the instrumental LSF for LAMOST, both the median residual and the robust residual scatter decrease. For $R^+_{8542}$, the median residual decreases from 0.013 to 0.007 and $\sigma_{\mathrm{MAD}}$ from 0.021 to 0.010, indicating improved consistency between the two surveys. A similar improvement is found for $R^+_{\mathrm{HK}}$, for which the median residual decreases from 0.017 to 0.004 and $\sigma_{\mathrm{MAD}}$ decreases from 0.023 to 0.012. The remaining median residual likely reflects survey-to-survey differences, including differences in the effective spectral resolution and other observational systematics, whereas the residual scatter may also include contributions from measurement noise and intrinsic activity variability between non-contemporaneous observations. Although the use of the LSF improves the agreement between surveys, systematically negative $R^+_{8542}$ values still persist.

\subsubsection{Propagation of Measurement Uncertainties}\label{subsubsec:Error}

Uncertainties in RV, SNR, atmospheric parameters, continuum normalization, and template interpolation propagate into the derived activity indices. For example, the radial velocity uncertainties of the LAMOST spectra have a median value of $\sim 3.8~\mathrm{km\,s^{-1}}$ according to the AFGK catalog. Such uncertainties shift the line center relative to the integration window and thus affect $R^{+}$.

\begin{figure}[htbp]
    \centering
    \includegraphics[width=0.95\linewidth]{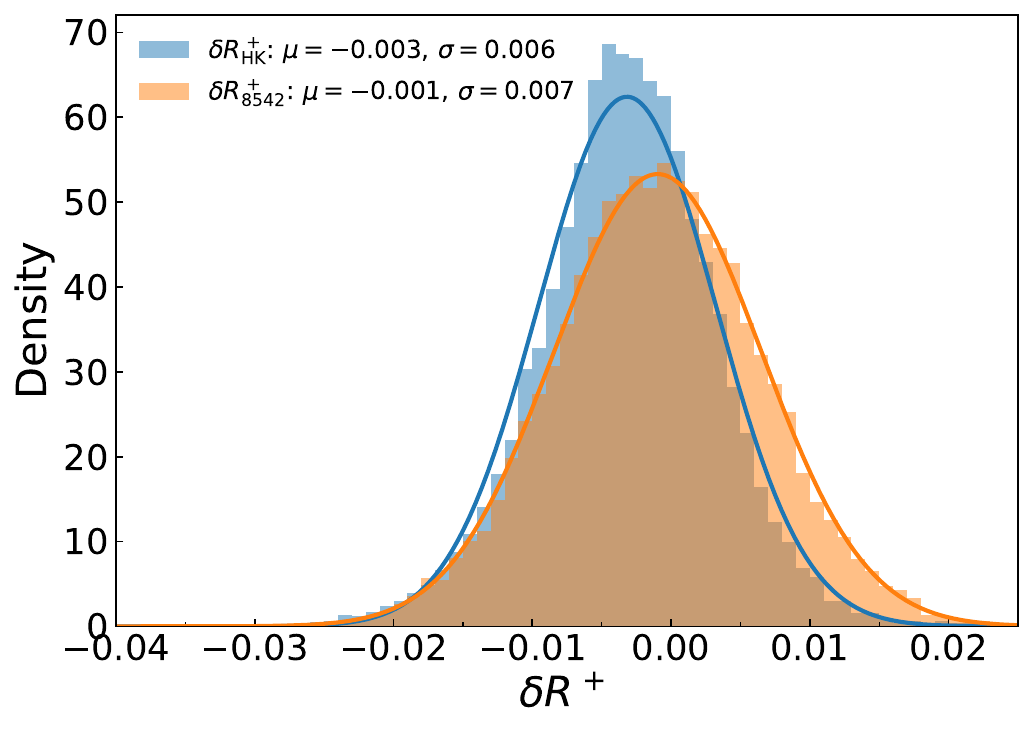}
    \caption{Distributions of the estimated uncertainties, $\delta R^{+}_{\rm HK}$ and $\delta R^{+}_{8542}$, derived from simulations based on the A1 templates. The Gaussian fits and their corresponding means and standard deviations are also shown.}
    \label{fig:errorlamost}
\end{figure}

\begin{figure*}[htbp] 
    \centering
    \begin{subfigure}{0.95\textwidth}
        \includegraphics[width=\textwidth]{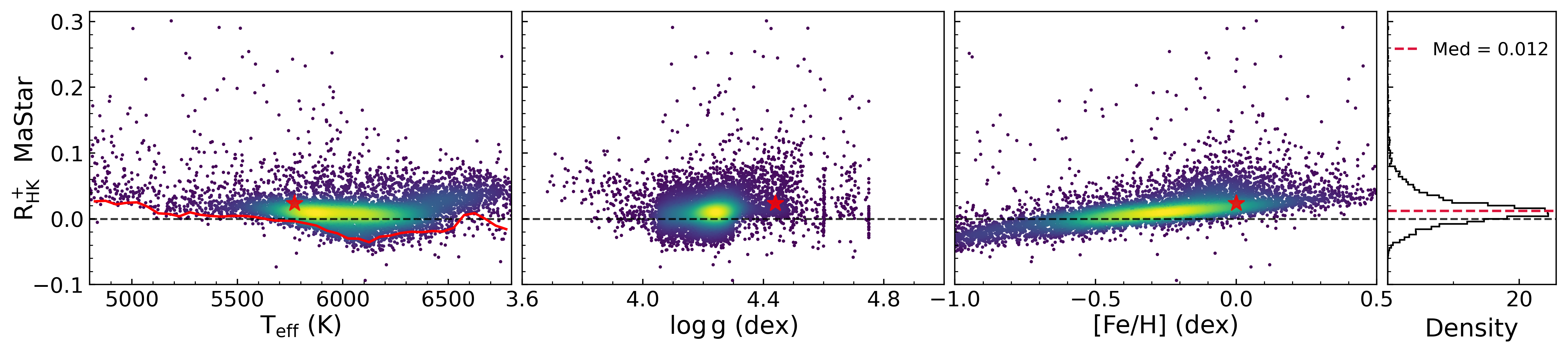}
    \end{subfigure}
    \begin{subfigure}{0.95\textwidth}
        \includegraphics[width=\textwidth]{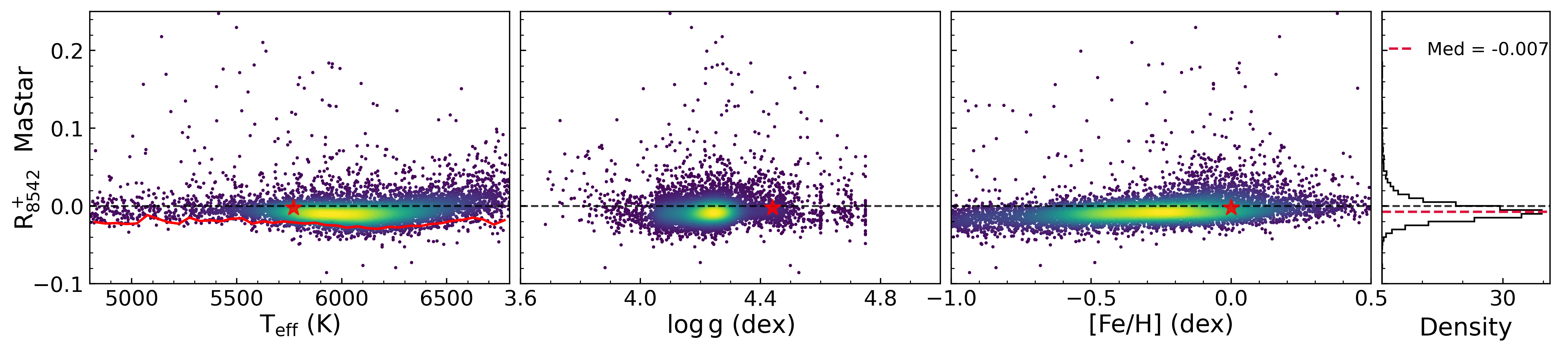}
    \end{subfigure}    
    
    \begin{subfigure}{0.95\textwidth}
        \includegraphics[width=\textwidth]{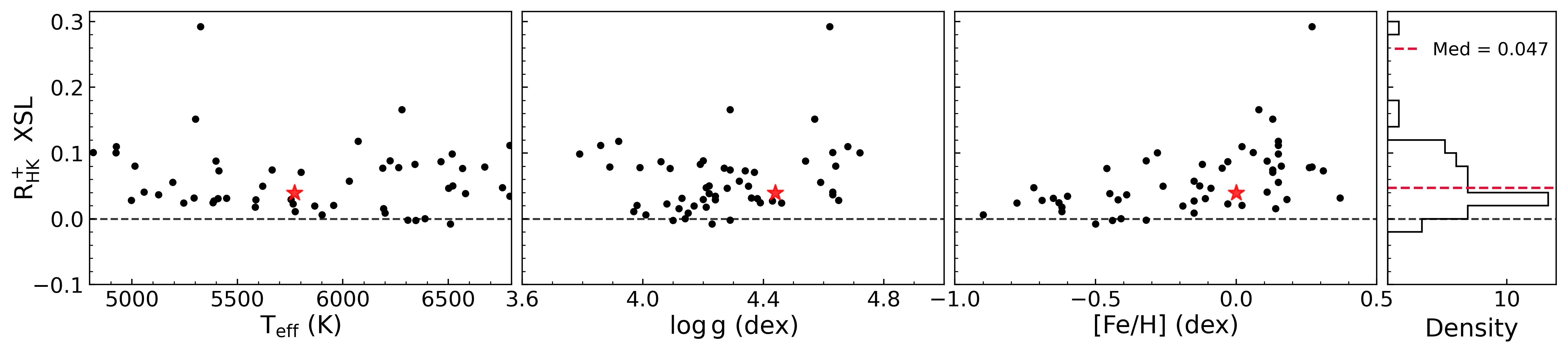}
    \end{subfigure}
    \begin{subfigure}{0.95\textwidth}
        \includegraphics[width=\textwidth]{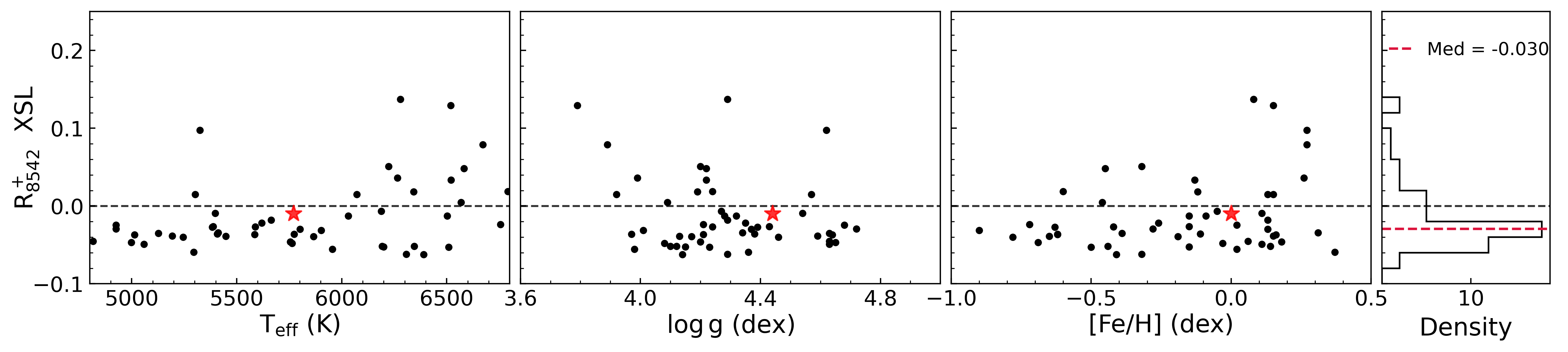}
    \end{subfigure}
    \caption{$R^+_{\rm HK}$ and $R^+_{8542}$ as functions of $T_{\rm eff}$, $\log g$, and [Fe/H] for the MaStar and XSL samples, derived using the A1 templates with $V_{\rm mic} = V^{\rm t}_{\rm mic}$. The upper two rows show the MaStar sample, and the lower two rows show the XSL sample. The red star marks the Sun. In the $T_{\rm eff}$ panels, the red curve denotes the 5th-percentile relation. High-density regions are shown in yellow, and the median lines are plotted in red. The black horizontal dashed line indicates $R^+ = 0$.}
    \label{fig:Rplusmastarandxsl}
\end{figure*}

To quantify the combined impact of these uncertainties, we performed simulations based on the A1 templates. Each template was first degraded to the LAMOST spectral resolution using a Gaussian kernel, and Gaussian noise consistent with the adopted SNR was then added. A Gaussian RV perturbation with $\sigma = 3.8~\mathrm{km\,s^{-1}}$ was subsequently applied, after which the spectrum was resampled onto the LAMOST wavelength grid. The resulting spectra were then treated as simulated LAMOST observations. Median uncertainties in $T_{\mathrm{eff}}$, $\log g$, [Fe/H], and [$\alpha$/Fe] were adopted from the AFGK catalog as functions of the corresponding stellar parameters, and these parameters were perturbed by adding Gaussian noise with standard deviations equal to the corresponding median uncertainties. After incorporating all sources of uncertainty, we generated a full set of simulated observations and measured the activity indices following the standard procedure. Because the templates do not exhibit line-core emission and thus more closely resemble inactive stellar spectra, these results can also be regarded as an estimate of the lower $R^{+}$ distribution.

Figure~\ref{fig:errorlamost} shows the resulting distributions. The uncertainties, $\delta R^{+}_{\mathrm{HK}}$ and $\delta R^{+}_{8542}$, follow approximately normal distributions with standard deviations of 0.006 and 0.007, respectively. Their mean offsets are $-0.003$ for $\delta R^{+}_{\mathrm{HK}}$ and $-0.001$ for $\delta R^{+}_{8542}$, implying that the combined uncertainty sources can introduce a small negative bias. However, this effect is insufficient to explain the systematically negative $R^{+}_{8542}$ values observed in the LAMOST sample (see Figure~\ref{fig:Rpluslamost}).

\subsubsection{MaStar and XSL $R^+$ Results}\label{subsubsec:results}

We measured $R^+_{\rm HK}$ and $R^+_{8542}$ indices for the MaStar and XSL samples using the A1 templates with Gaussian convolution, as shown in Figure~\ref{fig:Rplusmastarandxsl}. For MaStar, we additionally corrected $\log g$ and [Fe/H]. Although we did not perform an external comparison of the XSL atmospheric parameters, we consider them sufficiently reliable because the wings of the Ca\,\textsc{II} lines agree well with those of the A1 templates at the same adopted parameters.

For MaStar, most stars have $R^+_{\rm HK} > 0$, whereas more than half exhibit negative $R^+_{8542}$ values. The empirical lower envelope, $R^+_{8542,{\rm lower}}$, remains consistently below zero. The median value of $R^+_{8542}$ is approximately $-0.007$, which is lower than that of the LAMOST sample in Figure~\ref{fig:Rpluslamost} by 0.007. This offset is comparable to the systematic offset between LAMOST and MaStar discussed in Section~\ref{subsubsec:LSF}.
For XSL, only a few stars exhibit negative $R^+_{\rm HK}$ values, whereas the majority have $R^+_{8542} < 0$. The median $R^+_{8542}$ is even lower, reaching approximately $-0.03$. This stronger negative offset can be related to the higher spectral resolution of XSL, which preserves more of the line-core structure.


Overall, the recurrence of negative $R^+_{8542}$ values across these surveys indicates that observational systematics mainly contribute to scatter and small offsets, rather than to the full systematic negative bias.
\subsection{Evidence for Template-Related Mismatch}\label{subsec:temissue}


We therefore turn to the template side and compare synthetic spectra with the solar spectrum to identify the origin of the line-core mismatch.
\subsubsection{Solar $R^+$ Indices}

The Sun provides an important reference for interpreting the measured activity indices. We adopt the very high resolution ($\lambda/\Delta\lambda > 300{,}000$) optical and near-infrared solar flux atlas of \citet{Wallace2011}. Because this atlas represents a disk-integrated solar spectrum, it includes the effects of solar rotation, center-to-limb variation, and convection, and is therefore directly comparable to unresolved stellar spectra. The atlas is based on observations obtained in 1981 and 1989, corresponding to phases near the maximum of the solar activity cycle.

\begin{figure}[htbp]
    \centering
    \includegraphics[width=0.95\linewidth]{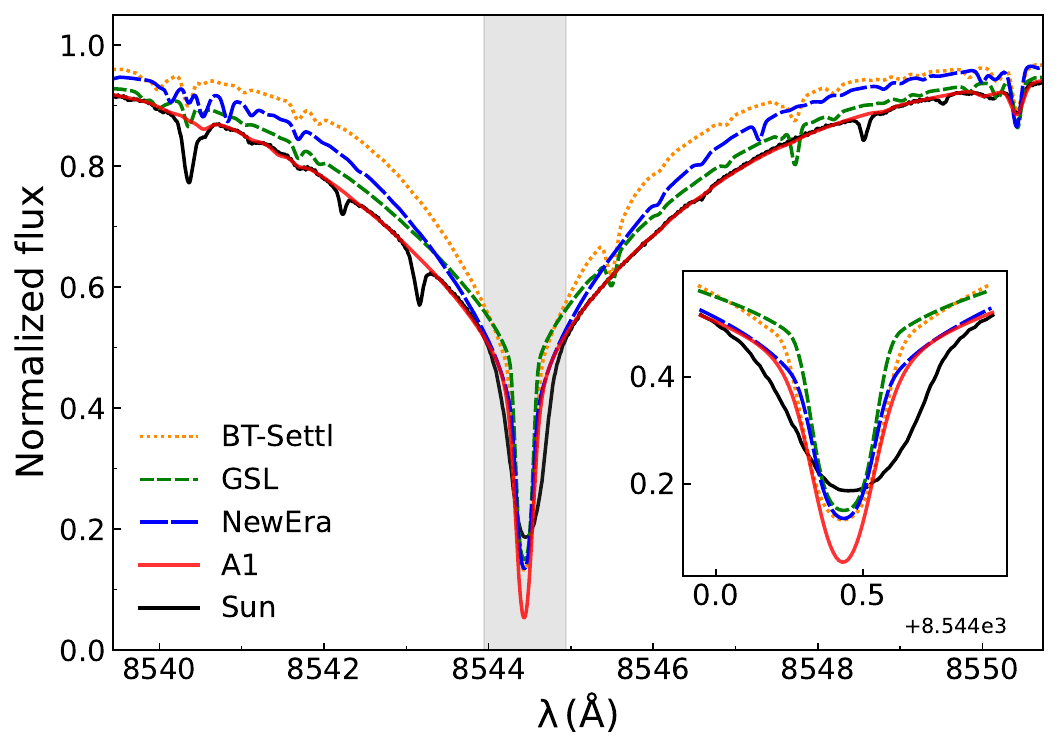}
    \caption{Comparison of the Ca\,\textsc{II} $\lambda8542$ line between the templates and the observed solar spectrum at solar parameters. The shaded region and the inset indicate the 1~\AA\ line-core window. The spectral resolving power is $\geq 300{,}000$, and solar rotational broadening ($v\sin i = 1.6~\mathrm{km\,s^{-1}}$) is included.}
    \label{fig:tem_sun}
\end{figure}

Its high spectral resolution allows the solar spectrum to be processed using the same analysis pipeline as the synthetic templates, thereby ensuring methodological consistency. We therefore compute the solar $R^+$ indices by convolving both the solar spectrum and the A1 template with a Gaussian kernel to match the spectral resolution of the corresponding surveys. The solar position is highlighted in Figures~\ref{fig:Rpluslamost} and~\ref{fig:Rplusmastarandxsl}, where the Sun is typically classified as a moderately active star \citep{Baliunas1995,Hall2007,Grijs2021}. At LAMOST resolution, the Sun has $R^+_{\rm HK}=0.0252$ and $R^+_{8542}=-0.0018$. The slightly negative value of $R^+_{8542}$ indicates that negative residuals are not restricted to inactive stars, but can also occur at solar-like activity levels. This comparison further shows that the template does not fully reproduce the observed solar line cores, particularly in the IRT.

\subsubsection{Line-Core Depth Discrepancy}\label{subsec:core_discrepancy}
 
Figure~\ref{fig:tem_sun} presents a comparison of the Ca~\textsc{II} $\lambda8542$ line between multiple template spectra and the solar spectrum. Although the A1 template performs best among the tested templates, reproducing the observed line wings while maintaining sufficiently deep line cores, its $\lambda8542$ line core remains slightly shallower than that of the observed solar spectrum. Integrating over the 1~\AA\ core window yields $R=0.370$ for the A1 template and $R=0.352$ for the Sun; the latter is smaller, resulting in negative $R^+_{8542}$ values. This direct high-resolution line-core measurement further supports the conclusion that negative IRT residuals can also occur at solar-like activity levels.
The asymmetry observed in the solar line core likely reflects the complex velocity structure of the chromosphere \citep{Uitenbroek2006,Landstreet2009}. This asymmetry can be caused by asymmetric vertical motions in the chromosphere, where upward and downward motions are not sampled equally. When averaged over time, this imbalance makes the blue side appear relatively higher than the red side of the line center.

\begin{figure*}[htbp]
    \centering
    \includegraphics[width=0.98\textwidth]{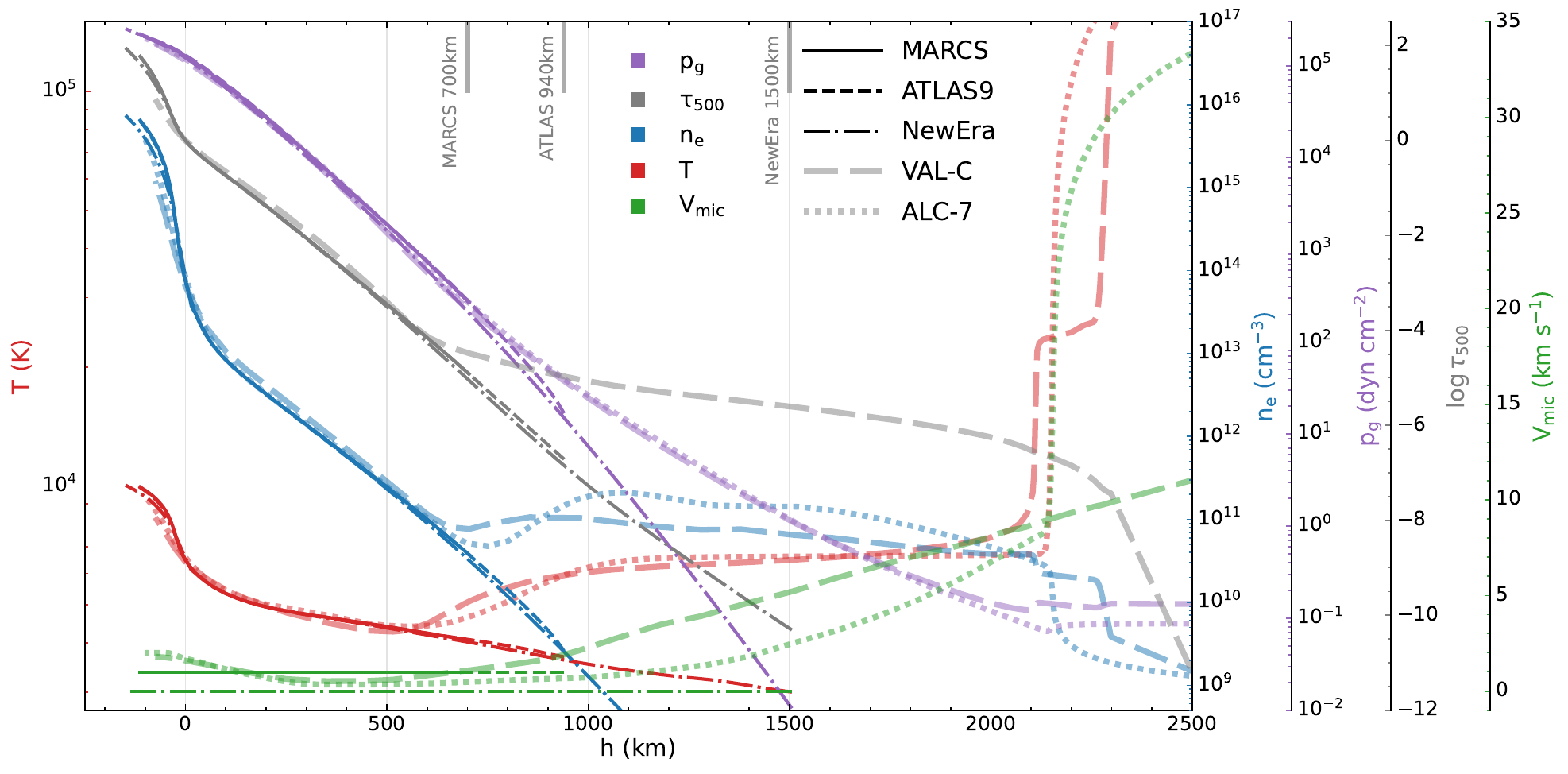}
    \caption{Solar atmospheric structures of the ATLAS9, MARCS, and NewEra model atmospheres, compared with those of the semi-empirical quiet-Sun models VAL-C and ALC-7. The displayed quantities, distinguished by color, are the gas pressure $p_{\rm g}$, continuum optical depth at 500~nm, $\tau_{500}$ (defined such that $\tau_{500}=1$ at height $h=0$), electron number density $n_{\rm e}$, temperature $T$, and microturbulent velocity $V_{\mathrm{mic}}$. Line styles distinguish the models, with the semi-empirical models shown as thicker and lighter lines. The thick gray vertical lines in the upper part indicate the approximate cutoff heights of the MARCS, ATLAS9, and NewEra models. At $h \approx 50$~km, $T \approx T_{\mathrm{eff}} = 5774$~K. For MARCS and ATLAS9, $V_{\mathrm{mic}} = 1~\mathrm{km\,s^{-1}}$; for NewEra, $V_{\mathrm{mic}} = 0~\mathrm{km\,s^{-1}}$.}
\label{fig:diffsolarmodel}
\end{figure*}

Similarly, the $\lambda8542$ core in the templates is shallower than that in the XSL spectra. This discrepancy is unlikely to be explained solely by the choice of convolution kernel or by resolution mismatch, as reproducing the XSL core depth would require degrading the templates to approximately twice the XSL resolving power ($\sim 20000$). Therefore, the discrepancy in the line cores likely has a more fundamental physical origin.
\subsubsection{Model Atmosphere Structure}\label{subsec:source}

Because different parts of a spectral line form at different atmospheric heights, the atmospheric structure---particularly in the upper photosphere and chromosphere---directly influences the emergent line core.
In Figure~\ref{fig:diffsolarmodel}, we compare the solar atmospheric structures of the MARCS, ATLAS9, and NewEra models with those of the semi-empirical quiet-Sun chromospheric models VAL-C \citep{Vernazza1981} and ALC-7 \citep{Avrett2008}.

To place the different models on a common geometrical-height scale, we define $h=0$ at $\tau_{500}=1$, consistent with the convention adopted for both semi-empirical models.
The MARCS solar structure\footnote{\url{https://marcs.astro.uu.se/data.html}} \citep{Gustafsson2008} provides a sufficiently complete set of atmospheric parameters, including the Rosseland optical depth $\tau_{\rm Ross}$, the monochromatic optical depth $\tau_{500}$, and the geometrical height $h$, with its original height scale referenced to $\tau_{\rm Ross}=1$.
We therefore shifted the MARCS height scale so that $h=0$ corresponds to $\tau_{500}=1$.
The quantities required for Figure~\ref{fig:diffsolarmodel} that are not directly tabulated in the model atmospheres were derived as described below.

The ATLAS9 solar model\footnote{\url{https://wwwuser.oats.inaf.it/fiorella.castelli/sun.html}; solar abundances from \citet{Asplund2005}} \citep{Castelli2003} does not tabulate $\tau_{\rm Ross}$, $\tau_{500}$, or $h$.
The Rosseland optical depth at each layer was calculated from the tabulated column mass $m$ and Rosseland mean opacity $\kappa_{\rm Ross}$ ($\mathrm{cm^2\,g^{-1}}$) by discretely integrating $d\tau_{\rm Ross}=\kappa_{\rm Ross}\,dm$.
We then estimated $\tau_{500}$ and $h$ using the relations among $\log\tau_{\rm Ross}$, $\log\tau_{500}$, and $h$ derived from the MARCS solar model.
Over the relevant atmospheric range, the relation between $\log\tau_{\rm Ross}$ and $\log\tau_{500}$ is well described by a linear fit, with a coefficient of determination of $R^{2}=0.99$ and a maximum absolute residual of 0.03~dex.
The relation between $\log\tau_{500}$ and $h$ is also approximately linear near both boundaries of the MARCS atmosphere (see Figure~\ref{fig:diffsolarmodel}), particularly in the outer layers, allowing modest linear extrapolation beyond the tabulated range.
This procedure yields an ATLAS9 height range of approximately $h\simeq-115$--$940$~km.
We also independently reconstructed the geometrical height scale from the model's column-mass and thermodynamic structure, assuming an ideal gas and neglecting molecular species.
This calculation gives a closely matching range of $h\simeq-110$--$962$~km, supporting the reliability of the adopted MARCS-based height calibration.

The NewEra solar model\footnote{\url{https://doi.org/10.25592/uhhfdm.16727}} \citep{Hauschildt2025} directly provides the geometrical radius, column mass, $\kappa_{\rm Ross}$, and electron pressure $P_{\rm e}$ at each atmospheric layer.
We calculated $\tau_{\rm Ross}$ by integrating $d\tau_{\rm Ross}=\kappa_{\rm Ross}\,dm$.
The resulting $\tau_{\rm Ross}$ values were converted to $\tau_{500}$ using the MARCS-based relation described above, after which the NewEra height scale was shifted so that $h=0$ at $\log\tau_{500}=0$.
The electron number density was calculated from the electron pressure according to
$n_{\rm e}=P_{\rm e}/(k_{\rm B}T)$, where $k_{\rm B}$ is the Boltzmann constant.

In the lower atmosphere, the temperature and electron-density stratifications of the theoretical models are broadly consistent with those of the semi-empirical models.
In the outer layers, however, the MARCS, ATLAS9, and NewEra atmospheres extend only to approximately $h\simeq700$, $940$, and $1500$~km, respectively.
Their uppermost layers are also cooler and have lower electron densities than the corresponding layers of the semi-empirical models.
Most importantly, they do not reproduce the chromospheric temperature rise, which begins at a height of approximately 500--600~km, as shown in Figure~\ref{fig:diffsolarmodel}.
Under the Eddington--Barbier approximation, the emergent line-center intensity is primarily determined by the layer at which $\tau_{\nu}\simeq1$.
In the quiet Sun, the line centers of the H\&K and IRT lines form at characteristic heights of approximately $h\simeq1900$ and $1450$~km, respectively \citep{Bjorgen2018,2012decs.confE.110R,Vernazza1981}.
These heights lie above the outer boundaries of the MARCS and ATLAS9 atmospheres and, for the H\&K lines, also above the outer boundary of the NewEra model.

For the solar Ca\,\textsc{II} lines modeled by \citet{Bjorgen2018}, the K-line-center optical depth remains as large as $\tau_{\nu}\sim10^{4}$ at $h\simeq700$~km, whereas the MARCS atmosphere terminates near this height.
Consequently, in spectral synthesis based on such a truncated structure, the H\&K and IRT line cores can be regarded as effectively forming in the uppermost layers of the photospheric atmosphere.

In the solar chromosphere, conventionally defined as extending from above the temperature minimum to the onset of the steep temperature rise into the transition region \citep{Linsky2017}, Ca\,\textsc{II} remains the dominant ionization stage of calcium throughout most of the line-forming region \citep{Carlsson2007,Wedemeyer2011}.
Its level populations do not generally follow the local LTE Saha--Boltzmann distribution. Moreover, nonradiative heating and dynamical processes can perturb these populations. For example, during the passage of shocks, the rising temperature redistributes Ca\,\textsc{II} from the ground state into excited levels.
In particular, the populations of the metastable $3d\,^2D$ levels, which are the lower levels of the IRT transitions, increase, whereas the population of the $4s\,^2S$ ground state, the lower level of the H\&K transitions, decreases but remains dominant \citep{Wedemeyer2011}.
The additional Ca\,\textsc{II} opacity associated with these chromospheric layers is absent from the MARCS atmosphere.
Consequently, stars with chromospheres can exhibit IRT absorption cores that are deeper than those predicted by synthetic photospheric templates.
The observed IRT cores therefore naturally yield negative $R^+_{\rm IRT}$ values after template subtraction in stars with low or moderate activity, as illustrated by the solar case.
Only at sufficiently high activity levels does chromospheric emission fill in the absorption and eventually drive the IRT line cores into emission \citep{Lanzafame2023,Linsky2017}.
This absorption-first behavior is analogous to the sequence described by \citet{Cram1979}, who showed that adding chromospheric material to model M-dwarf atmospheres can initially deepen Balmer absorption lines before driving them into emission.

Moreover, the contrasting behavior of H\&K and the IRT, with $R^+_{\rm HK}$ usually being positive and $R^+_{\rm IRT}$ often negative, can be understood in terms of their different source-function behavior.
In the solar chromosphere, the H\&K lines generally exhibit a stronger source-function enhancement \citep{2012decs.confE.110R,Bjorgen2018}.
In particular, at the heights where $\tau_\nu\simeq1$ for the K$_2$ and H$_2$ features, the source function exceeds its value at the upper boundary of the photospheric model, so these features tend to appear in emission. By contrast, at the $\tau_\nu\simeq1$ formation heights of the IRT cores, the source function remains below its value at the upper boundary of the photosphere, so the line cores appear darker than the radiation emerging from the photosphere.
In K- and M-type stars, the photospheric continuum flux around H\&K is lower than that around the IRT. A modest chromospheric activity flux can therefore produce a larger line-to-continuum contrast in H\&K, making the chromospheric contribution more readily detectable as positive residual emission, whereas a substantially larger chromospheric flux is required to produce a comparable effect in the IRT \citep{Lanzafame2023}.
\added{Because our sample is dominated by F- and G-type stars and contains relatively few K-type stars, we expect on qualitative grounds that the continuum-flux effect plays a less prominent role in the contrasting behavior of $R^+_{\rm HK}$ and $R^+_{\rm IRT}$ than it would in cooler-star samples.}

\subsubsection{NLTE Effects}\label{subsec:NLTE}

Non-local thermodynamic equilibrium (NLTE) effects can become important in the chromosphere because of its lower densities and the presence of a temperature inversion relative to the photosphere. While the line wings are generally close to LTE, NLTE effects primarily influence the line cores. Their impact also becomes stronger at lower metallicity, leading to a reduced line-core flux relative to the LTE case and thus producing a larger negative residual \citep{Andretta2005,Mashonkina2007}. Because observed spectra naturally include these NLTE effects whereas LTE templates do not, this mismatch can contribute to cases with $R^+ < 0$ when LTE-based templates are used.

\begin{figure}[htbp]
    \centering
    \includegraphics[width=0.98\linewidth]{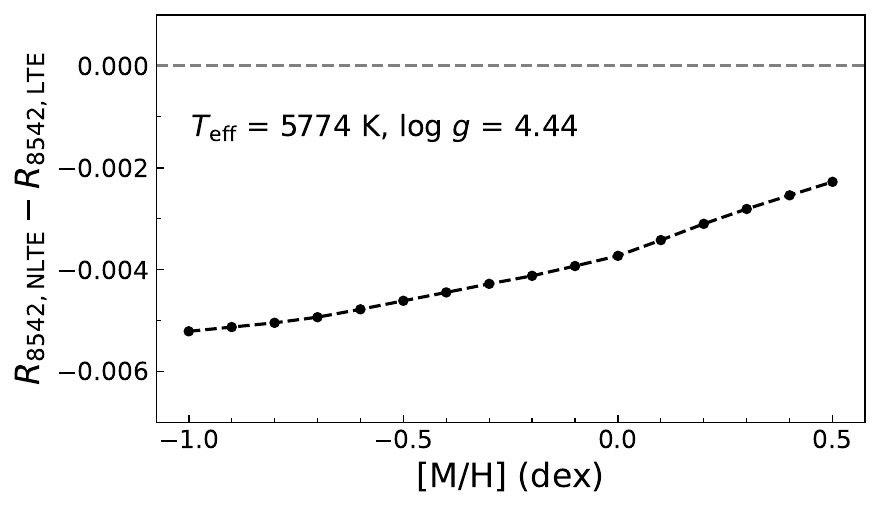}
    \caption{Difference between the NLTE and LTE Ca~\textsc{II} $\lambda8542$ line-core $R$ indices as a function of [M/H], computed at solar $T_{\rm eff}$ and $\log g$ with high spectral resolution.}
    \label{fig:NLTEvsLTE}
\end{figure}

If LTE templates are adopted, $R^+$ would in principle be expected to increase with [Fe/H]. Such a trend is seen for the H\&K lines, but not for the IRT lines (see Figures~\ref{fig:Rpluslamost} and~\ref{fig:Rplusmastarandxsl}), likely because the IRT cores form lower in the atmosphere and are therefore less sensitive to NLTE effects. 
To further quantify the possible contribution of NLTE effects, we computed the Ca~\textsc{II} $\lambda8542$ line-core index ($R_{8542}$) in both LTE and NLTE at solar atmospheric parameters while varying the global metallicity $[\mathrm{M/H}]$, using spectra synthesized with PySME \citep{JianPySMEv1,PySME2023} at high spectral resolution. The MARCS model atmospheres, VALD line list, and Grevesse2007 solar abundance scale were adopted. The Ca NLTE treatment is based on the departure-coefficient grids of \citet{Amarsi2020}. As shown in Figure~\ref{fig:NLTEvsLTE}, the NLTE treatment yields a deeper line core than in the LTE case, producing a negative offset. The magnitude of this offset increases toward lower $[\mathrm{M/H}]$, but remains only at the level of a few $\times 10^{-3}$; at $[\mathrm{M/H}]$  = 0, it is about $-0.004$, substantially smaller in magnitude than the observed solar offset of approximately $-0.02$ discussed in Section~\ref{subsec:core_discrepancy}. This indicates that NLTE effects alone cannot account for the full discrepancy.
We therefore conclude that NLTE effects likely contribute to the offset, but the dominant source is more likely to be the absence of chromospheric structure in the photospheric models.

\section{An Empirical Mitigation Based on Microturbulence}\label{sec:vmic_cor}
To explore whether the observed--template line-core mismatch can be reduced within the framework of photospheric templates, we test an empirical adjustment to the adopted microturbulent velocity.
Macroturbulence mainly broadens absorption-line profiles while approximately conserving their equivalent widths, and is therefore not expected by itself to remove the systematically negative residual core fluxes.
Typical photospheric microturbulent velocities derived from weak lines are $V_{\mathrm{mic}}^{\rm t}\sim 1$--2~km\,s$^{-1}$ for FGK stars \citep{Mucciarelli2011,Gray4th,Gaiabenchmark2014}.
For weak lines, $V_{\mathrm{mic}}$ mainly broadens the profile with little change in equivalent width, whereas for saturated lines it can strengthen the absorption and alter the curve of growth \citep{Gray4th}.
We therefore increase $V_{\mathrm{mic}}$ as an empirical way to deepen the synthetic Ca~\textsc{II} IRT cores and compensate for the missing physics in photospheric templates.
This adjustment should be regarded as an empirical correction rather than a measurement of the true microturbulent velocity field.

\begin{figure*}[htbp]
    \centering
    \begin{subfigure}[b]{0.46\textwidth}
        \includegraphics[width=\textwidth]{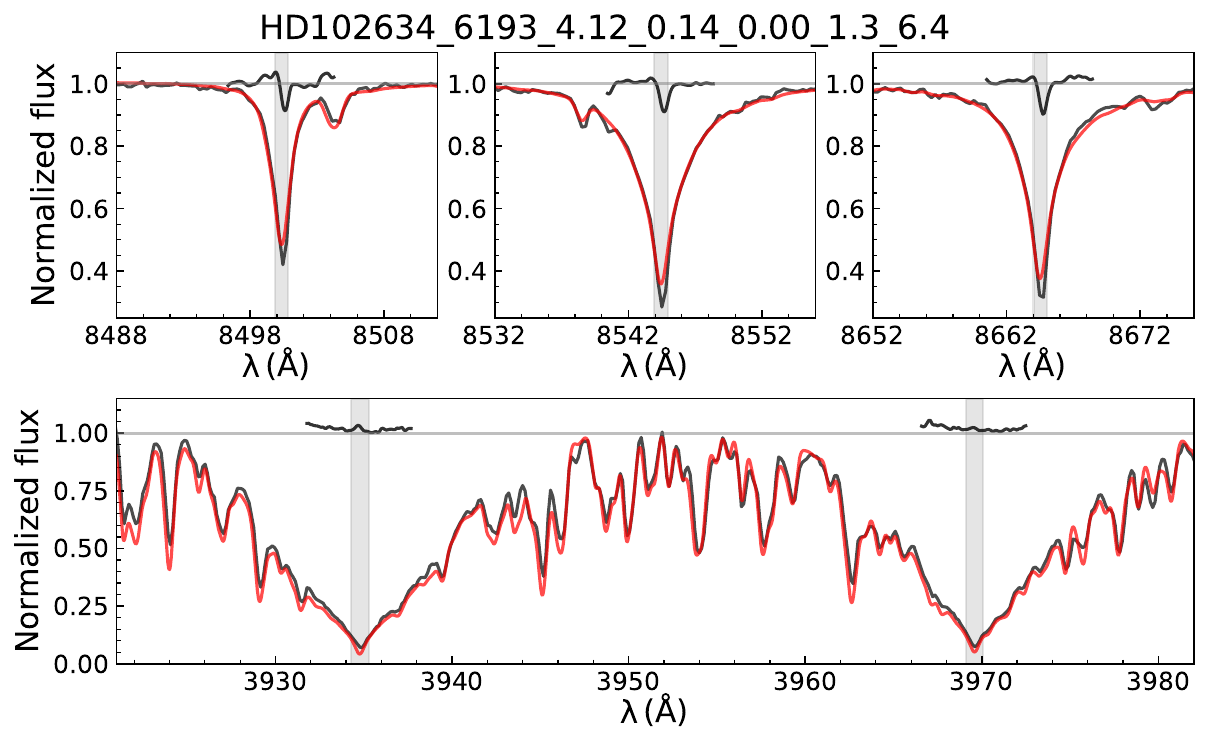}
    \end{subfigure}
    \hspace{0.0\textwidth}
    \begin{subfigure}[b]{0.46\textwidth}
        \includegraphics[width=\textwidth]{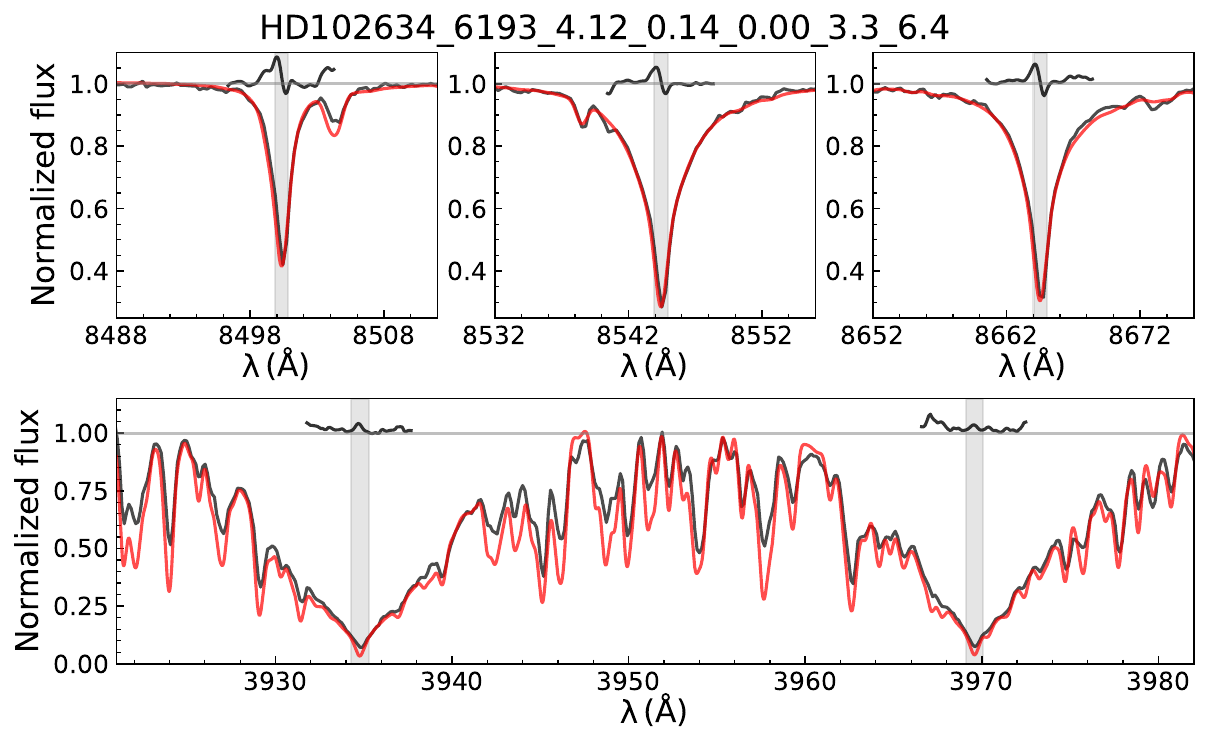}
    \end{subfigure}
    \begin{subfigure}[b]{0.46\textwidth}
        \includegraphics[width=\textwidth]{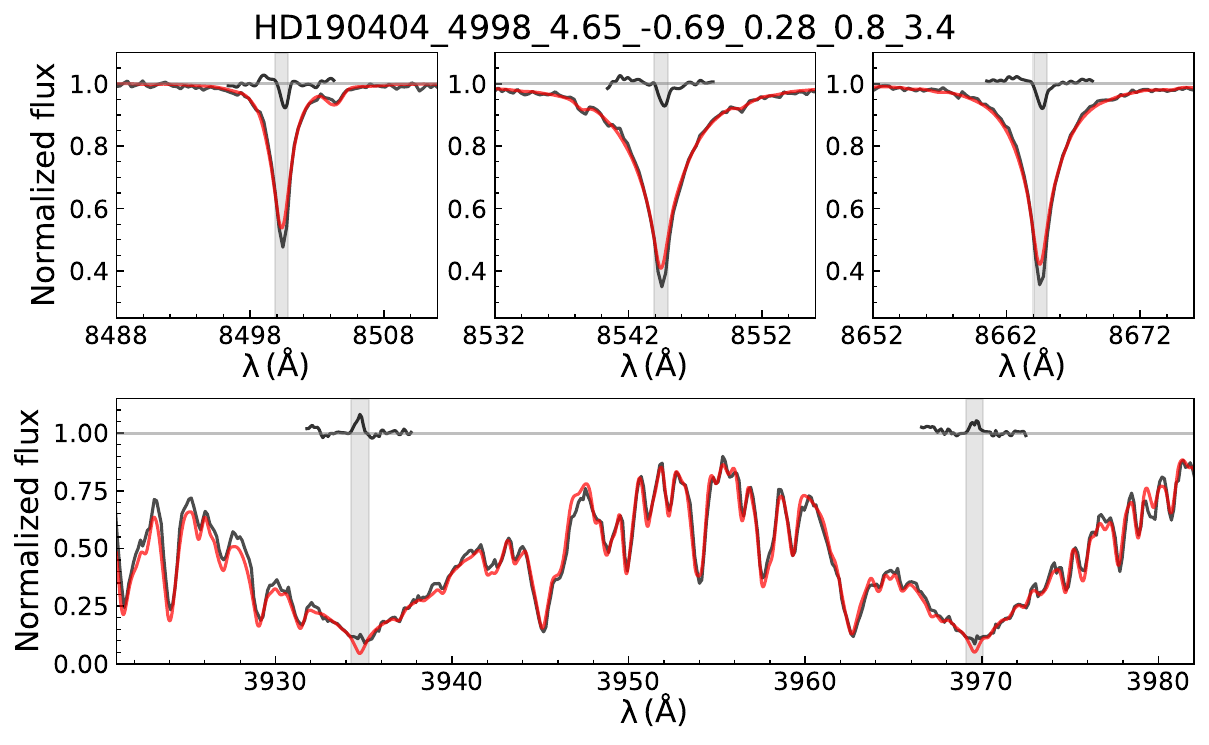}
    \end{subfigure}
    \hspace{0.0\textwidth}
    \begin{subfigure}[b]{0.46\textwidth}
        \includegraphics[width=\textwidth]{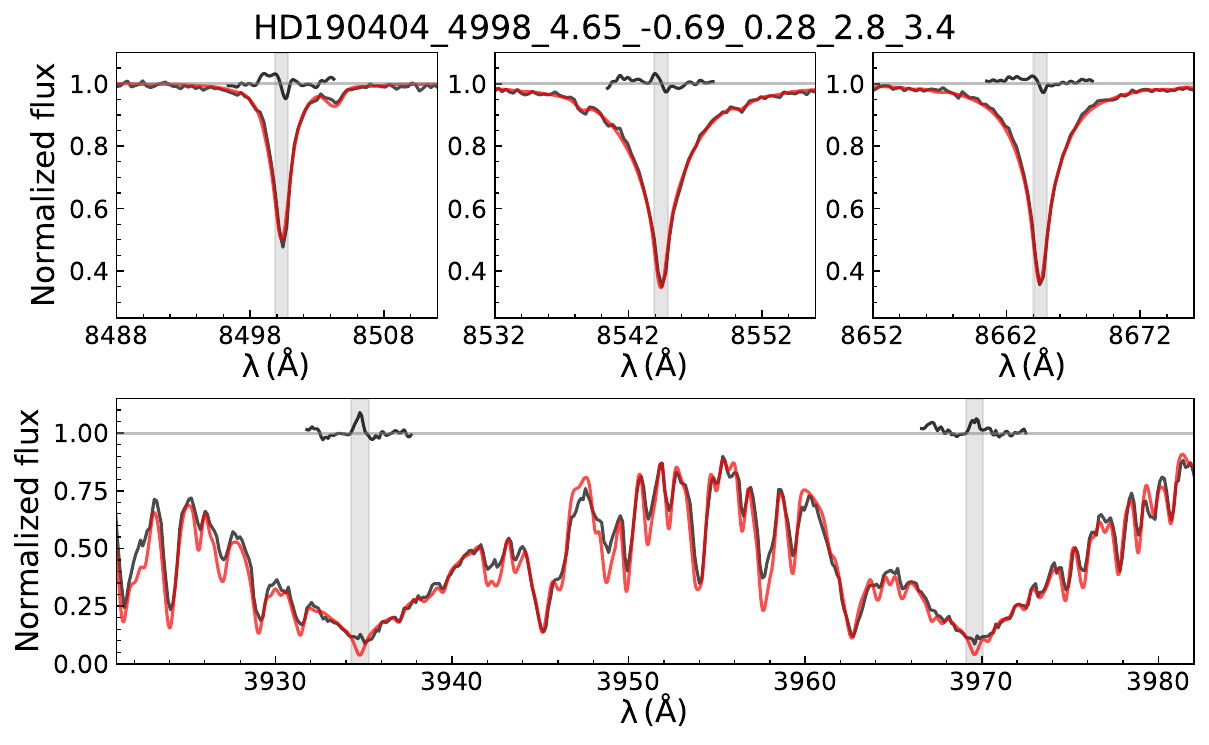}
    \end{subfigure}
    \caption{Comparison of the Ca\,\textsc{II} H\&K and IRT line profiles between XSL spectra (black) and A1 templates (red). The left-hand panels show $V_{\mathrm{mic}} = V^{\rm t}_{\mathrm{mic}}$, and the right-hand panels show $V_{\mathrm{mic}} = V^{\rm t}_{\mathrm{mic}} + 2~\mathrm{km\,s^{-1}}$. The upper panels display the hot, metal-rich star HD~102634, while the lower panels display the cool, metal-poor star HD~190404; neither star exhibits obvious emission cores. The panel titles list the stellar name, $T_{\rm eff}$, $\log g$, [Fe/H], [$\alpha$/Fe], and the adopted $V_{\mathrm{mic}}$ and $V_{\mathrm{mac}}$. The residual spectra are vertically shifted to unity for clarity. The gray shaded regions mark the 1~\AA\ line-core windows. The templates are convolved to the XSL resolving power of $\lambda/\Delta\lambda = 9774$.}
    \label{fig:xsl_Caline}
\end{figure*}

\begin{figure*}[htbp]
    \centering
    \includegraphics[width=0.98\textwidth]{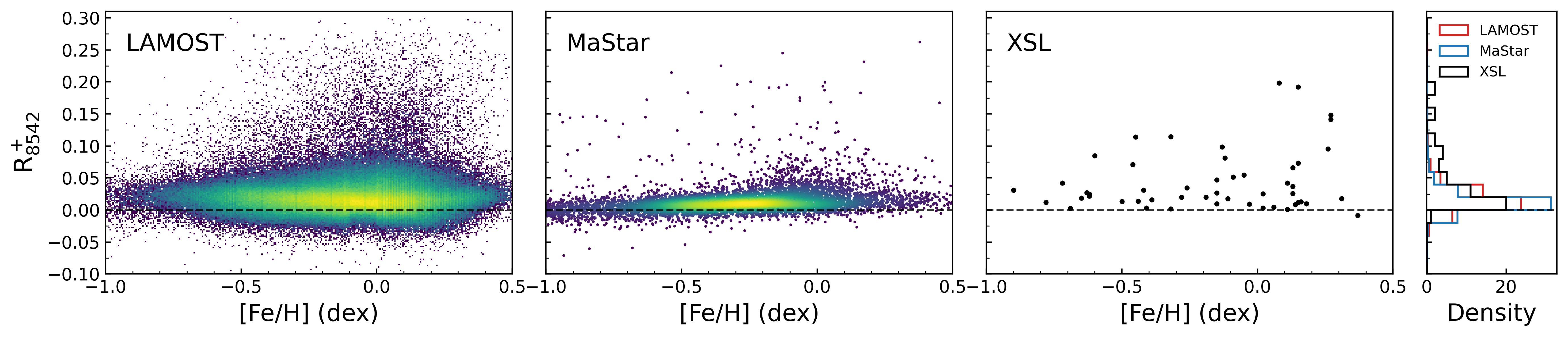}
    \caption{$R^+_{8542}$ as a function of [Fe/H] for LAMOST, MaStar, and XSL when using A1 templates with $V_{\mathrm{mic}} = V^{\mathrm{t}}_{\mathrm{mic}} + 2$~km\,s$^{-1}$.}
    \label{fig:Rplus8542vmic+2}
\end{figure*}

Using \texttt{iSpec} with the A1 configuration, we compared templates computed with different $V_{\mathrm{mic}}$ values against the XSL spectra of HD~102634 and HD~190404 in Figure~\ref{fig:xsl_Caline}.
When $V_{\mathrm{mic}} = V^{\rm t}_{\mathrm{mic}} + 2$~km\,s$^{-1}$ is adopted in the right-hand panels, the synthetic IRT cores become deeper and provide a substantially better match to the observed cores than in the left-hand panels, where the default choice $V_{\mathrm{mic}} = V^{\rm t}_{\mathrm{mic}}$ is used. We therefore apply this empirical adjustment to the full solar-like XSL sample in this work. As shown in Figure~\ref{fig:Rplus8542vmic+2}, this adjustment shifts nearly all XSL $R^+_{8542}$ values above zero.

We then recomputed $R^+_{\rm HK}$ and $R^+_{8542}$ for the LAMOST and MaStar samples using $V_{\mathrm{mic}} = V^{\rm t}_{\mathrm{mic}} + 2$~km\,s$^{-1}$. 
With this empirical correction, the median $R^+_{8542}$ increases by 0.016 for LAMOST and by 0.010 for MaStar, moving both medians above zero (see Figure~\ref{fig:Rplus8542vmic+2}). 
By contrast, $R^+_{\rm HK}$ decreases by 0.009 for LAMOST and by 0.008 for MaStar. 
Under the present index definition, this behavior is expected because increasing $V_{\mathrm{mic}}$ also deepens the many neighboring absorption lines in the H\&K region; after degradation to low resolution, these changes lower the local pseudo-continuum level and therefore affect $R^+_{\rm HK}$. 
This empirical correction is therefore mainly applicable to the IRT cores of solar-like FGK stars, but is less suitable for H\&K.

We emphasize that the increased $V_{\mathrm{mic}}$ is not intended to represent the true photospheric microturbulent velocity, nor does it provide a physical explanation for the observed--template mismatch. Instead, it is an empirical tuning parameter used to deepen the synthetic IRT cores and reduce the zero-point offset within the A1 photospheric-template framework.
Its required value is therefore expected to depend on the adopted synthesis setup, stellar-parameter range, and spectral line considered. In this work, the choice of $V_{\mathrm{mic}} = V_{\mathrm{mic}}^{\rm t}+2~\mathrm{km\,s^{-1}}$ should be regarded only as an A1-specific empirical mitigation, not as a physically motivated solution.

\begin{figure*}[htbp]
    \centering
    \includegraphics[width=0.95\textwidth]{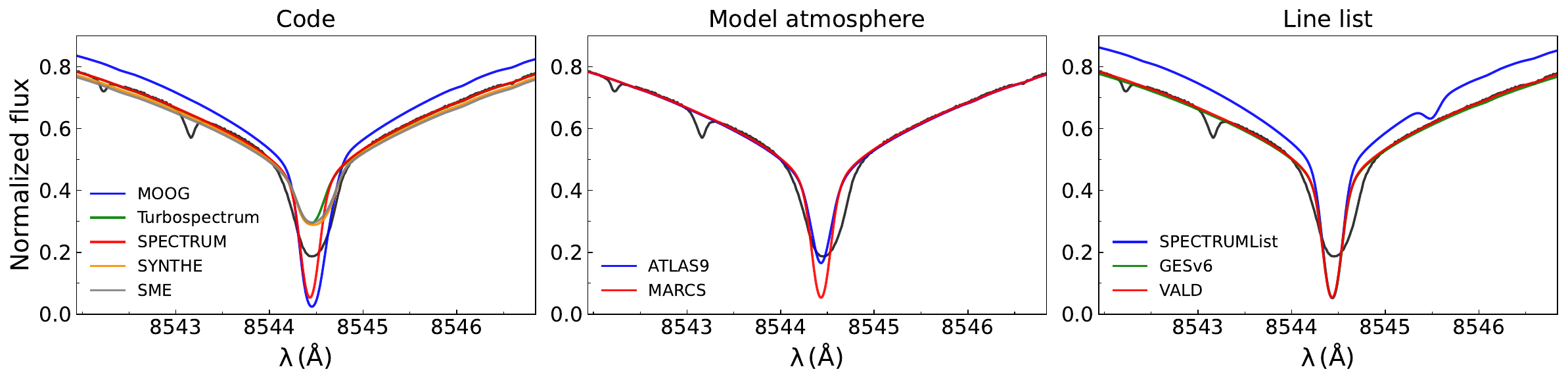}
    \caption{Effects of different radiative transfer codes, model atmospheres, and line lists on the Ca\,\textsc{II} $\lambda8542$ profile under solar parameters. The black curve shows the observed solar spectrum. The reference \texttt{iSpec} configuration adopts SPECTRUM, MARCS, and VALD, and only one component is varied at a time. The solar abundances are adopted from Grevesse2007.}
    \label{fig:ispec_factor}
\end{figure*}

\section{Template Dependence and Cross-Calibration of $R^+$}\label{sec:consistency}

Because the derived $R^+$ values depend on the adopted photospheric templates, we examine how the synthesis configuration affects the measured indices and the extent to which results from different template sets can be cross-calibrated. We first compare the effects of key model components on the IRT line profiles and use this comparison to motivate our fiducial template configuration. 
We then quantify the configuration dependence of $R^+_{\mathrm{HK}}$ and $R^+_{8542}$ across a broader set of synthetic spectra, and finally assess how these differences affect the inferred lower envelope and the interpretation of survey-based activity measurements.

\subsection{Adoption of the Fiducial Template Configuration}\label{subsec:temcompare}

We used \texttt{iSpec} to examine how the main synthesis components---including the radiative transfer code, model atmosphere, and atomic line list---affect the Ca~\textsc{II} IRT profiles. 
For this purpose, we compared synthetic spectra with the observed solar spectrum of \citet{Wallace2011}. 
Because the differences introduced by the Grevesse2007 and Asplund2009 solar abundance scales are small in this comparison, we do not discuss them separately here.

As shown in Figure~\ref{fig:ispec_factor}, the choice of radiative transfer code produces noticeable differences in the profile of Ca~\textsc{II} $\lambda8542$. 
Among SPECTRUM, MOOG, Turbospectrum, SYNTHE, and SME, MOOG yields the deepest line core but the shallowest wings, whereas spectra synthesized with SPECTRUM show a relatively deeper line core and better agreement with the observed solar spectrum. Changing the model atmosphere from MARCS to ATLAS9 also affects the core depth, with the MARCS models producing deeper cores. 
We further compared three line lists---SPECTRUMList, GESv6, and VALD---and found that the line-core differences are small, while the wing differences are more evident, with SPECTRUMList producing shallower wings.

On the basis of this comparison, we adopt the A1 configuration as the fiducial template set used throughout the remainder of this section. 
This configuration combines the SPECTRUM radiative transfer code, MARCS model atmospheres, the Grevesse2007 solar abundances, and the VALD line list. 
Among the tested combinations, it provides the most satisfactory compromise between reproducing the observed line wings and maintaining sufficiently deep line cores.

\subsection{Inter-comparison of Synthetic Template Sets}\label{subsec:template_intercompare}

Different configurations produce different synthetic spectra. 
For survey applications, it is therefore important to quantify how strongly the derived $R^+$ indices depend on the adopted synthesis configuration, and whether the resulting differences are dominated by systematic offsets. To this end, we generated 40 sets of synthetic spectra. 
Each configuration is labeled as A1, B1, C1, etc.\ (see Table~\ref{tab:config_mapping_separated}). 
Because the GESv6 line list does not cover the H\&K region, only 20 configurations yield $R^+_{\mathrm{HK}}$. 
In addition, because the ATLAS9-based templates provide limited coverage in $[\alpha/\mathrm{Fe}]$, we restrict this comparison to stars with $0 \leq [\alpha/\mathrm{Fe}] \leq 0.4$.

Using the LAMOST spectra and the LSF-convolved templates, we measured $R^+_{\mathrm{HK}}$ and $R^+_{8542}$ for all tested configurations. The full numerical comparisons relative to A1, including the correlation coefficient ($r$), the linear-fit parameters (slope and intercept), the mean residual (bias), and the residual dispersion ($\sigma$), are provided in Appendix~\ref{app:consistency_full}. Overall, $R^+_{\mathrm{HK}}$ behaves similarly to $R^+_{8542}$, and we therefore focus primarily on $R^+_{8542}$ in the discussion below. 

\begin{figure*}[htbp]
    \centering
    \begin{subfigure}{0.95\textwidth}
        \includegraphics[width=\textwidth]{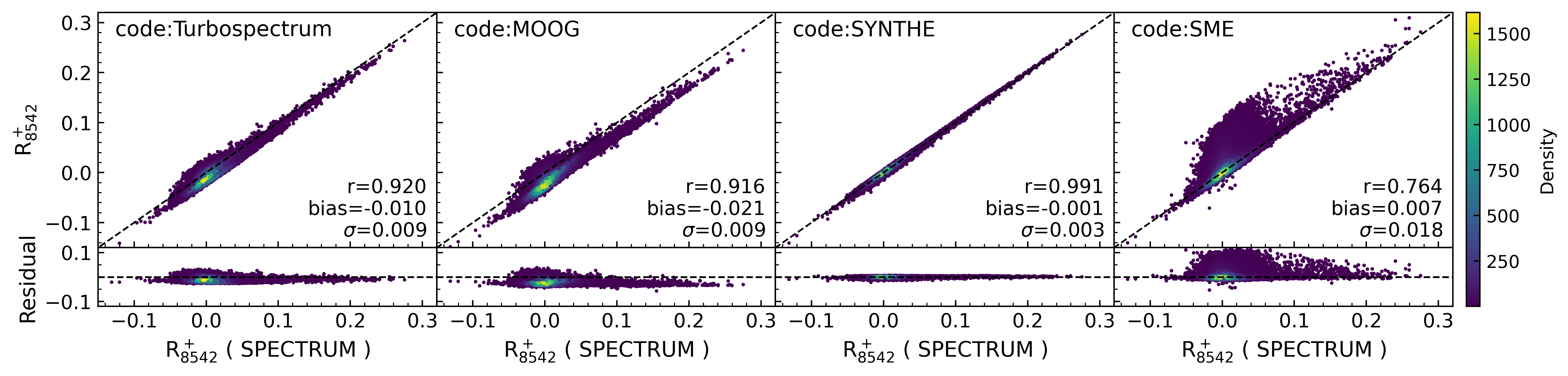}
    \end{subfigure}
    \begin{subfigure}{0.75\textwidth}
        \includegraphics[width=\textwidth]{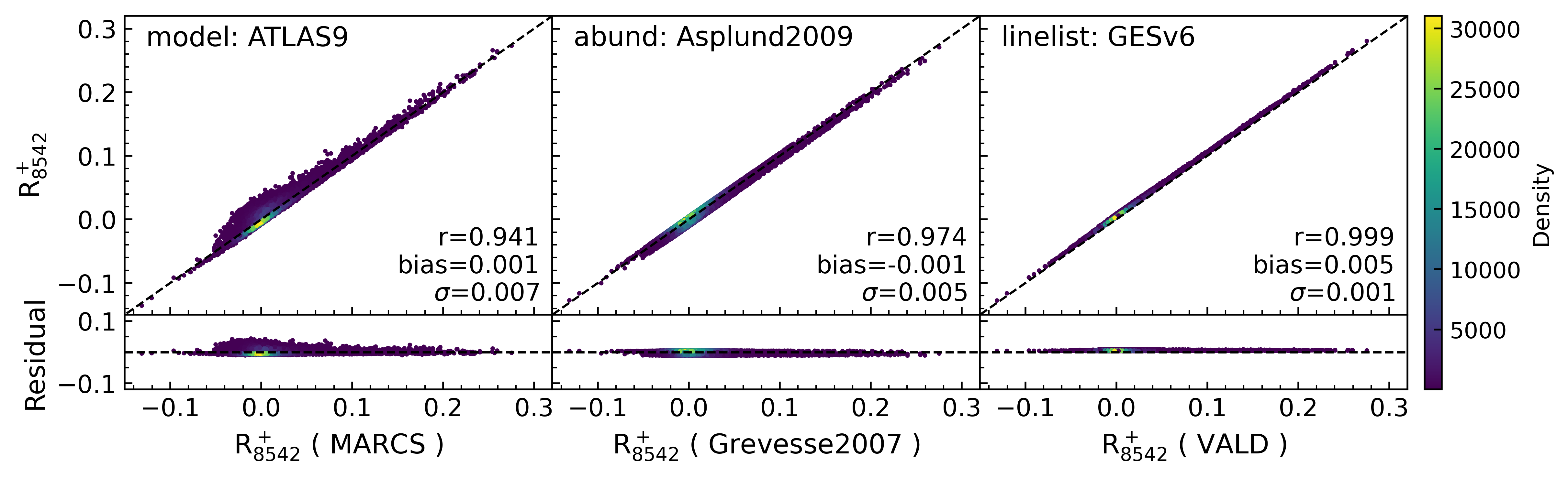}
    \end{subfigure}
    \begin{subfigure}{0.53\textwidth}
        \includegraphics[width=\textwidth]{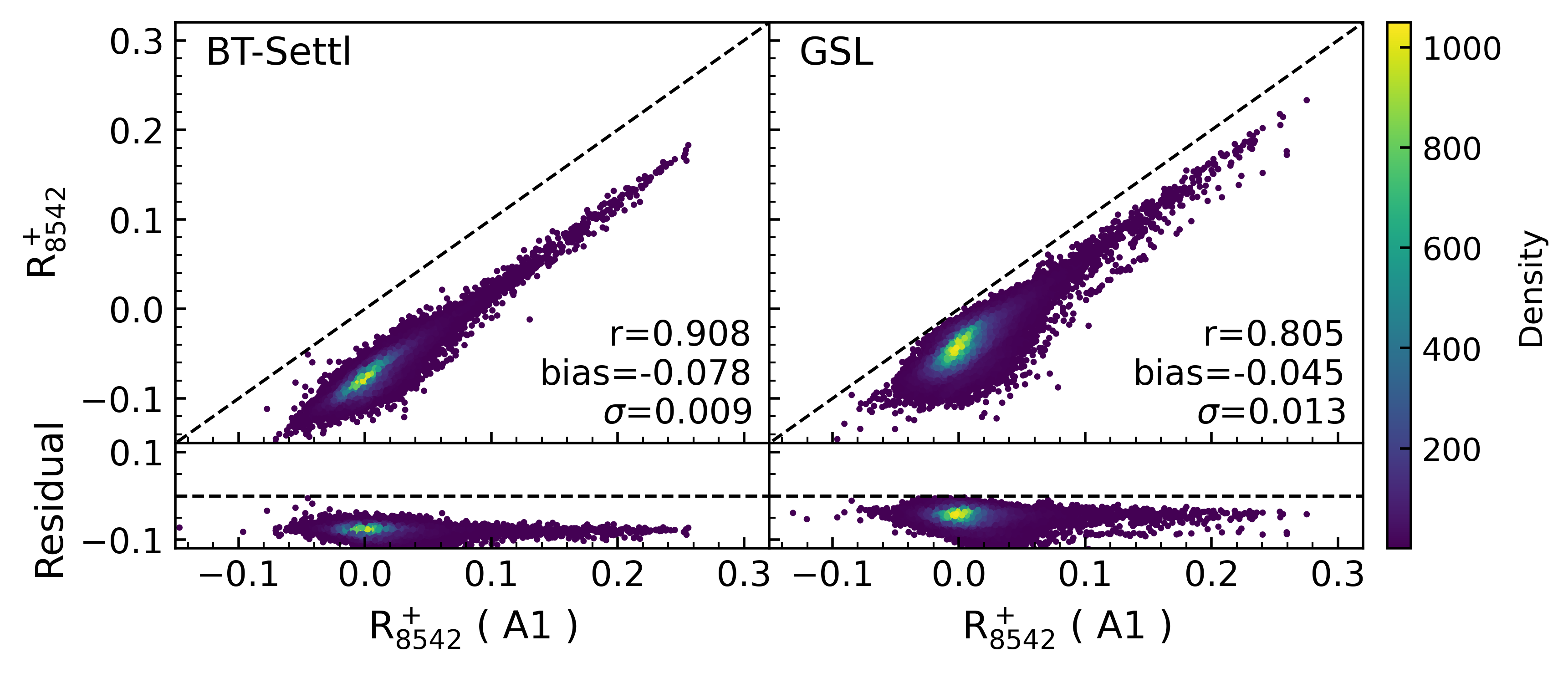}
    \end{subfigure}
    \caption{Correlations and residuals of $R^+_{8542}$ for the LAMOST sample derived using different synthesis configurations, all evaluated relative to the A1 templates (x-axis). In each case, only one component of the A1 configuration is changed in order to isolate the effect of that specific choice; the modified component is indicated in the upper-left corner of the first- and second-row panels. The bottom row shows the results obtained with the BT-Settl and GSL libraries. The quantities $r$, bias, and $\sigma$ denote the correlation coefficient, mean residual, and standard deviation of the residuals, respectively.}
    \label{fig:diff_comb8542}
\end{figure*}

As listed in Table~\ref{tab:appendix_rplus_full}, linear fits relative to A1 yield slopes within $1 \pm 0.05$ for all configurations except those involving SME, for which the slopes remain within $1 \pm 0.1$ (configurations I and J). This pattern suggests that the differences among configurations are more likely driven by systematic offsets than by substantial changes in slope. 
This result has a practical implication: although the absolute zero point of $R^+$ is template-dependent, the activity indices derived from different synthetic template libraries remain strongly linearly correlated. Thus, $R^+$ measurements based on different template choices can be placed on a common scale through empirical calibration relations.
Figure~\ref{fig:diff_comb8542} compares the $R^+_{8542}$ values derived from different synthesis configurations with those obtained from the A1 templates and summarizes their agreement through the annotated values of $r$, bias, and $\sigma$. Strong positive correlations are preserved across most configurations. 
In the top panel, changing the radiative transfer code yields correlation coefficients above 0.9 for $R^+_{8542}$ in all cases except SME ($r = 0.764$). 

Among the radiative transfer codes, the offsets in $R^+_{8542}$ relative to SPECTRUM range from $\mathrm{bias} = -0.001$ for SYNTHE to $\mathrm{bias} = -0.021$ for MOOG, while the corresponding dispersions range from $\sigma = 0.003$ for SYNTHE to $\sigma = 0.018$ for SME. 
These results indicate that the radiative transfer code is the dominant source of configuration-dependent variation in the derived indices. 
By comparison, changing the model atmosphere from MARCS to ATLAS9 introduces a smaller offset ($\mathrm{bias} = 0.001$) and a moderate dispersion ($\sigma = 0.007$), indicating a secondary dependence on the model atmosphere.
The solar abundance scale and the line list produce comparatively small biases and dispersions, implying a weaker influence on $R^+_{8542}$.

In the bottom panel, we also compare the results obtained with the public BT-Settl and GSL libraries.
As summarized in Table~\ref{tab:appendix_rplus_full}, BT-Settl shows the largest systematic offset in $R^+_{8542}$ ($\mathrm{bias} = -0.078$), whereas GSL exhibits a relatively large dispersion ($\sigma = 0.013$). 
For the GSL library, part of the observed dispersion can be attributed to a discontinuity in the indices near $T_{\mathrm{eff}} \approx 5100$~K, as discussed in Appendix~\ref{app:libissues}.

\subsection{Lower Envelope Systematics}\label{subsec:model_dependence}

\begin{figure*}[htbp]
    \centering
    \begin{subfigure}[b]{0.9\textwidth}
        \includegraphics[width=\textwidth]{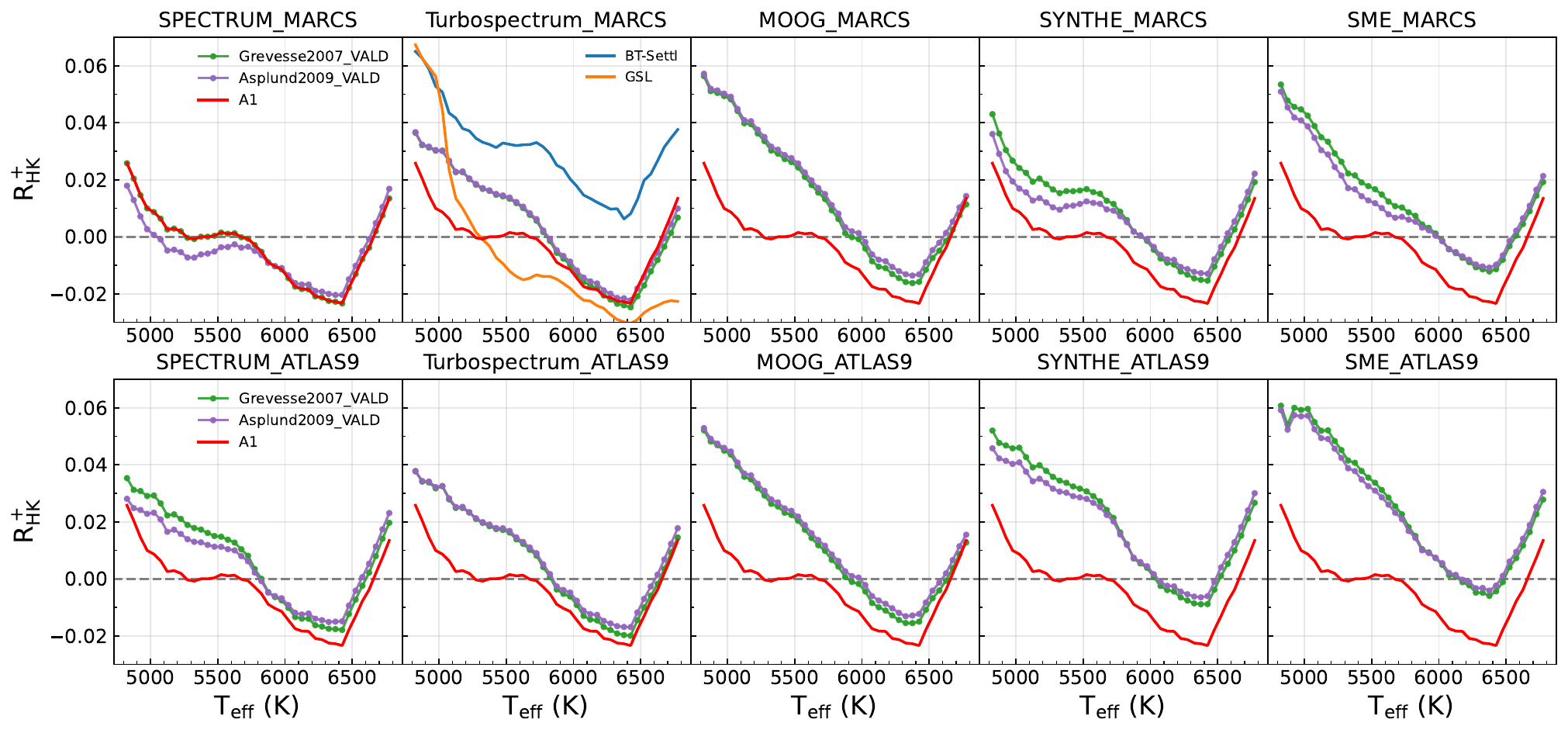}
    \end{subfigure}
    \begin{subfigure}[b]{0.9\textwidth}
        \includegraphics[width=\textwidth]{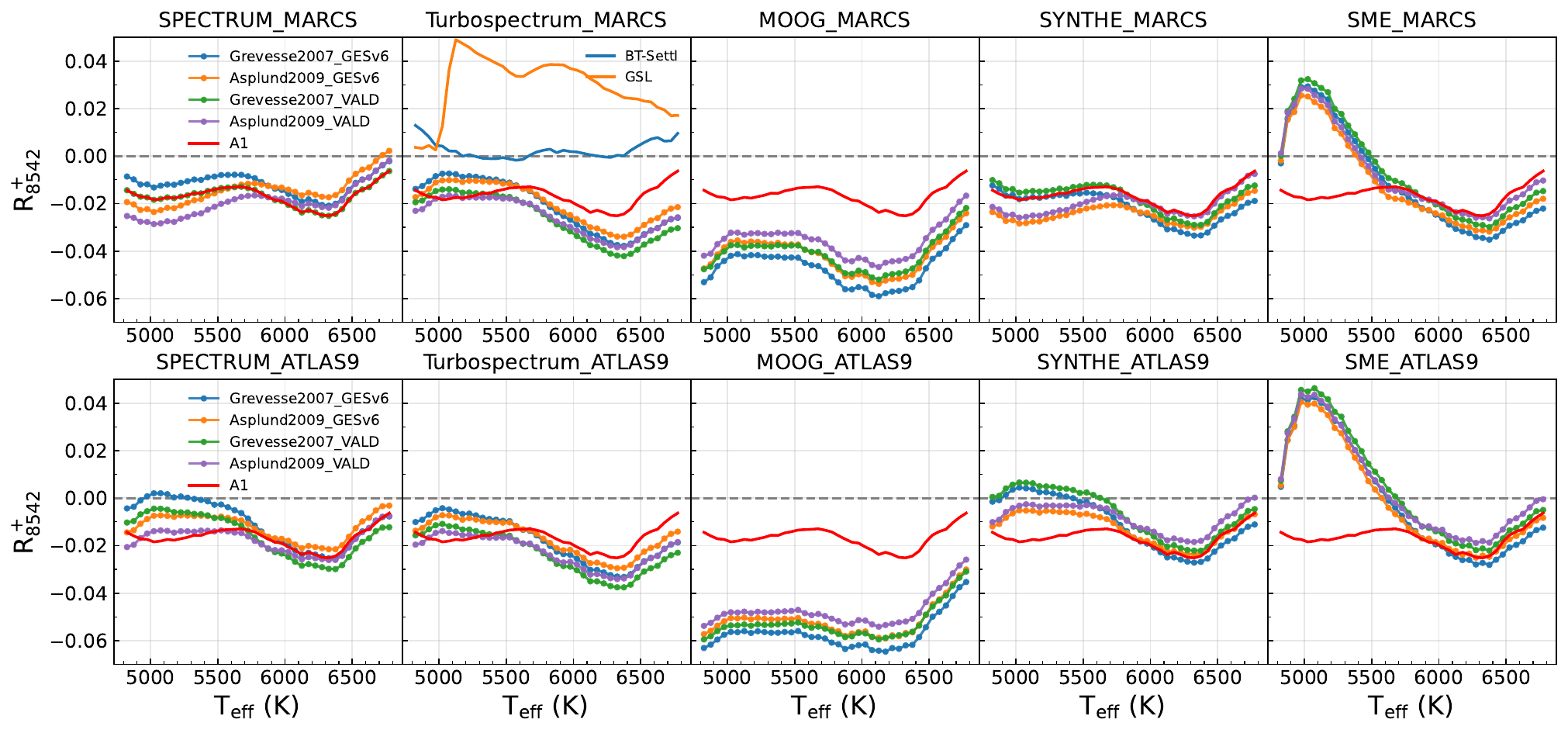}
    \end{subfigure}
    \caption{5th-percentile curves of $R^+_{\rm HK}$ and $R^+_{8542}$ for the LAMOST solar-like sample. The upper two rows show $R^+_{\rm HK}$, and the lower two rows show $R^+_{8542}$. In each panel, the title lists the radiative transfer code and model atmosphere, while the colors distinguish the combinations of solar abundance scale and line list. Each curve corresponds to a different synthesis configuration. The A1 5th-percentile curve is shown as a red solid line and is included in all panels for reference. For visual clarity, the BT-Settl and GSL curves are shifted upward by 0.025 in $R^+_{\rm HK}$ and by 0.1 in $R^+_{8542}$, respectively.}
    \label{fig:5thcurve}
\end{figure*}

Figure~\ref{fig:5thcurve} shows the 5th-percentile curves of $R^+_{\mathrm{HK}}$ and $R^+_{8542}$ as functions of $T_{\mathrm{eff}}$, which we use as empirical representations of the lower envelope. 
For clarity, the synthesis-configuration labels defined in Table~\ref{tab:config_mapping_separated} are not used in this figure.

The temperature dependence of the 5th-percentile curves is broadly similar across most synthesis configurations, but their absolute levels show clear systematic offsets. 
The largest variations again arise from the choice of radiative transfer code. 
By contrast, the 5th-percentile curves based on the VALD and GESv6 line lists exhibit similar temperature-dependent trends, as seen in the SPECTRUM\_MARCS panel for $R^+_{8542}$ in Figure~\ref{fig:5thcurve}. Compared to Grevesse2007, the curves based on Asplund2009 are higher at higher temperatures and lower at lower temperatures, but the overall trends remain similar.
These comparisons indicate that the solar abundance scale and line list have relatively minor effects on the inferred lower envelope.

The influence of the model atmosphere is more apparent at lower effective temperatures and depends on the radiative transfer code. 
For the SPECTRUM and SYNTHE cases, the MARCS-based 5th-percentile curves show a concave morphology at low temperatures, whereas the corresponding ATLAS9-based curves are more convex. 
For Turbospectrum, MOOG, and SME, the low-temperature curves are predominantly convex. This effect is particularly pronounced for SME, whose strong upward curvature at low temperatures contributes to the larger dispersion seen in Figure~\ref{fig:diff_comb8542}. 
At higher effective temperatures, the different configurations converge to broadly similar trends.

The public libraries show comparable behavior. 
In Figure~\ref{fig:5thcurve}, the BT-Settl curves for both $R^+_{\mathrm{HK}}$ and $R^+_{8542}$ are similar to those obtained with the SPECTRUM\_MARCS and SYNTHE\_MARCS configurations, whereas the GSL curves exhibit a discontinuity near $T_{\mathrm{eff}} \approx 5100$~K, consistent with the behavior discussed in Appendix~\ref{app:libissues}.

\subsection{Survey Implications}\label{subsec:survey_implications}

The results above have several implications for survey-based applications of template-subtracted Ca~\textsc{II} activity indices. 
First, the absolute scale of $R^+$ is not universal, but depends on the adopted synthesis configuration. 
As a result, measurements derived from different template sets should not be combined directly without accounting for systematic offsets. 

Second, the dominant source of configuration-dependent variation is the radiative transfer code, followed by the model atmosphere, while the solar abundance scale and line list have comparatively smaller effects. 
Maintaining a consistent synthesis setup is therefore important when comparing activity indices within or across large spectroscopic surveys.

Third, despite the systematic offsets, most configurations preserve strong linear correlations with the fiducial A1 results. This supports the practical use of template-subtracted $R^+$ diagnostics even when different theoretical template libraries are employed. This behavior indicates that cross-calibration onto a common reference system is feasible, provided that the calibration relation is established explicitly. Such a calibration can be constructed using the fitted slopes and intercepts reported in Appendix~\ref{app:consistency_full}. 



\section{Conclusions}\label{sec:conclusion}
We present a comprehensive investigation of the negative $R^+_{8542}$ values arising from template-subtraction measurements of Ca\,\textsc{II} activity indices. 
Using multiple synthetic spectral libraries, we measure $R^+_{\mathrm{HK}}$ and $R^+_{8542}$ for the LAMOST DR9, MaStar, and XSL DR3 samples, and assess the origin of the negative IRT residual indices and their implications for survey-based chromospheric activity measurements. 
Our main conclusions are as follows:

\begin{enumerate}

\item 

Observational effects alone do not explain the widespread negative $R^+_{8542}$ values. Stellar-parameter offsets, instrumental LSF treatment, and propagated measurement uncertainties mainly broaden the $R^+$ distributions and introduce small systematic shifts, but remain insufficient within realistic uncertainty ranges. The recurrence of negative $R^+_{8542}$ values in LAMOST, MaStar, and XSL indicates that the effect is not confined to a single survey or pipeline.


\item
Our results suggest that the frequent negative $R^+_{\rm IRT}$ values arise primarily from a mismatch between chromospheric stellar spectra and purely photospheric templates. In the chromosphere, Ca\,\textsc{II} remains the dominant ionization stage of calcium, while shock heating can increase the relative populations of its metastable $3d\,^2D$ levels; the resulting enhancement of Ca\,\textsc{II} IRT opacity relative to the photosphere allows the observed IRT cores to become deeper than the synthetic photospheric profiles in low- and moderately active stars. By contrast, the stronger source-function enhancement of the H\&K cores more readily produces positive residual emission, whereas the IRT cores can remain absorption-dominated until substantially higher activity levels are reached. 
\added{The lower near-UV continuum around H\&K may further enhance this contrast in cool K- and M-type stars, but we expect on qualitative grounds that this effect plays a smaller role in our sample, which is dominated by F- and G-type stars.
} NLTE effects contribute in the same direction, but their estimated magnitude is too small to dominate the observed offset.


\item 
An empirical increase in the adopted microturbulent velocity can partially mitigate the IRT core mismatch within the framework of photospheric templates. This adjustment shifts the $R^+_{8542}$ distributions of the survey samples toward positive values, but it is less suitable for $R^+_{\mathrm{HK}}$ under the present implementation. We therefore regard it as a practical empirical mitigation rather than a physically motivated solution.

\item 
The absolute scale of $R^+$ depends on the adopted synthesis configuration. The largest variations arise from the radiative transfer code, followed by the model atmosphere, while the solar abundance scale and line list have comparatively smaller effects.
At the same time, different configurations preserve strong linear correlations, indicating that $R^+$ measurements derived from different template sets remain useful for diagnosing chromospheric activity and can be cross-calibrated onto a common scale through empirical calibration relations.

\end{enumerate}

Overall, $R^+$ remains a useful tracer of chromospheric activity in large spectroscopic surveys, provided that the synthesis assumptions are applied consistently and that configuration-dependent systematic offsets are properly taken into account.
In particular, negative Ca\,\textsc{II} IRT residual indices should not be interpreted literally as negative chromospheric emission, but rather as evidence of a mismatch between the observed line core and the adopted photospheric template.


\begin{acknowledgments}

This work was supported by the National Natural Science Foundation of China (12273056, 12090041) and the National Key R\&D Program of China (2022YFA1603002). H.L.Y. and Z.R.B. were supported by the National Key R\&D Program of China (Grant No. 2023YFA1607901). X.F. acknowledges the support of the National Natural Science Foundation of China (NSFC) No. 12573040, No. 12533008, and No. 12203100.

We thank Kefeng Tan for helpful comments.

Guoshoujing Telescope (the Large Sky Area Multi-Object Fiber Spectroscopic Telescope LAMOST) is a National Major Scientific Project built by the Chinese Academy of Sciences. Funding for the project has been provided by the National Development and Reform Commission. LAMOST is operated and managed by the National Astronomical Observatories, Chinese Academy of Sciences.

This work made use of the MaNGA Stellar Library (MaStar), as released in SDSS DR17 (\url{https://www.sdss4.org/dr17/mastar}), and the X-shooter Spectral Library (XSL) Data Release 3 (\url{https://xsl.astro.unistra.fr}). We acknowledge the use of \texttt{iSpec}, PySME, and the VALD database for atomic and molecular line data.

\end{acknowledgments}

\facilities{LAMOST \citep{2012RAA....12.1197C}}

\software{\texttt{iSpec}, PySME, Astropy \citep{astropy:2013, astropy:2018, astropy:2022}, NumPy \citep{harris2020array}, Matplotlib \citep{Hunter:2007}, SciPy \citep{2020SciPy-NMeth}}




\appendix
\restartappendixnumbering


\section{Discontinuities in $R$ Values from the GSL and BT-Settl}\label{app:libissues}

\begin{figure*}[htbp]
    \centering
    \begin{subfigure}[t]{0.45\textwidth}
        \includegraphics[width=\textwidth]{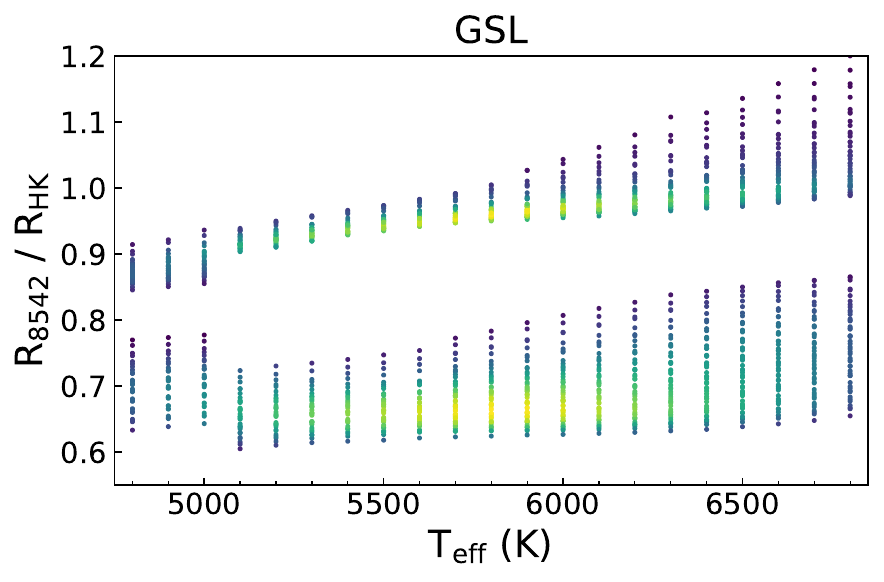}
    \end{subfigure}
    \hspace{0.02\textwidth}
    \begin{subfigure}[t]{0.45\textwidth}
        \includegraphics[width=\textwidth]{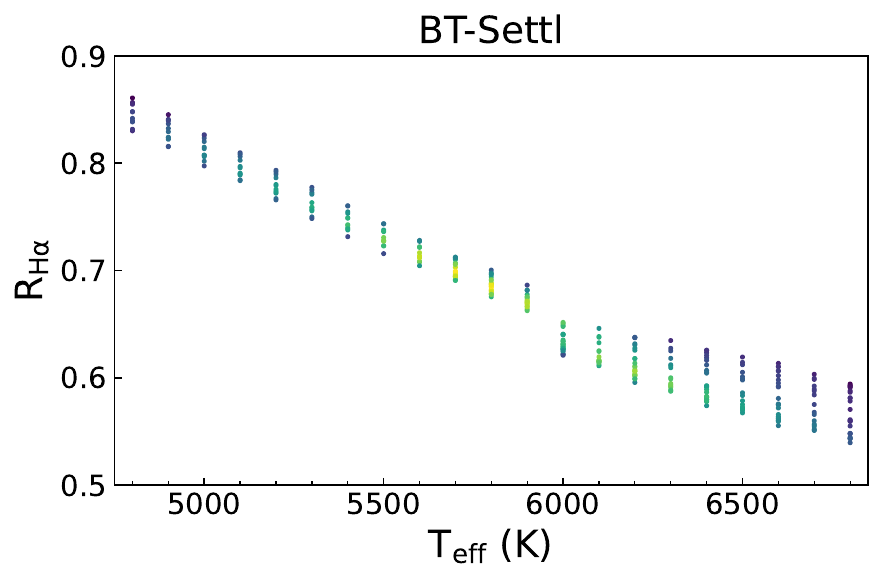}
    \end{subfigure}
    \vspace{-0.2cm}
    \caption{Discontinuities in the GSL and BT-Settl libraries. The left panel shows the GSL results, with the discontinuity occurring at $T_{\rm eff} \approx 5100$~K; $R_{\rm HK}$ is shifted upward for visual clarity, and $R_{8542}$ is shown in the lower part of the same panel. The right panel shows $R_{\mathrm{H}\alpha}$ from BT-Settl, with the discontinuity occurring at $T_{\rm eff} \approx 5900$~K. High-density regions are shown in yellow.}
    \label{fig:GSLBTissue}
\end{figure*}

We measured the $R$ values using the GSL and BT-Settl libraries. As shown in Figure~\ref{fig:GSLBTissue}, the $R$ values derived from the GSL exhibit an abrupt discontinuity around $T_{\mathrm{eff}} \approx 5100$~K, characterized by a sudden increase in the H\&K indices and a simultaneous decrease in the IRT indices with increasing temperature. A similar feature is seen in the H$\alpha$ line of the BT-Settl library around $T_{\mathrm{eff}} \approx 5900$~K.

\section{Full Numerical Results of the Consistency Tests}\label{app:consistency_full}

For completeness, we present in this appendix the full numerical results of the consistency tests for the LAMOST sample. Figure~16 in the main text already summarizes the overall behavior of the different synthesis configurations through the annotated values of $r$, bias, and $\sigma$.
Here we list the complete numerical results, including the fitted slope and intercept of the linear relations relative to the A1 templates, to facilitate direct comparison and reuse.

In the table below, the letter part of each label denotes the combination of radiative transfer code and model atmosphere, and the numeric suffix denotes the abundance scale and line list (see Table~\ref{tab:config_mapping_separated}). 
The linear fits are written in the form
\begin{equation}
y = {\rm slope}\times x + {\rm intercept},
\end{equation}
where $x$ and $y$ denote the indices derived from the A1 template set and the comparison configuration, respectively.

\begin{table*}[htbp]
\centering
\caption{Full numerical comparison of $R^+_{\rm HK}$ and $R^+_{8542}$ relative to the A1 templates for the LAMOST sample.}
\label{tab:appendix_rplus_full}
\renewcommand{\arraystretch}{0.85}
\setlength{\tabcolsep}{3pt}
\begin{tabularx}{\textwidth}{
>{\raggedright\arraybackslash}p{1.87cm}
*{5}{>{\centering\arraybackslash}X}
|*{5}{>{\centering\arraybackslash}X}
}
\hline
\hline
& \multicolumn{5}{c|}{$R^+_{\rm HK}$} & \multicolumn{5}{c}{$R^+_{8542}$} \\
Label & $r$ & bias & $\sigma$ & slope & intercept & $r$ & bias & $\sigma$ & slope & intercept \\
\hline
A1 & 1.000 & 0.000 & 0.000 & 1.000 & 0.000 & 1.000 & 0.000 & 0.000 & 1.000 & 0.000 \\
A2 & --    & --    & --    & --    & --    & 0.999 & 0.005 & 0.001 & 1.004 & 0.005 \\
A3 & 0.996 & $-$0.001 & 0.003 & 0.976 & $-$0.001 & 0.974 & $-$0.001 & 0.005 & 0.980 & $-$0.001 \\
A4 & --    & --    & --    & --    & --    & 0.981 & 0.004 & 0.004 & 0.984 & 0.004 \\
B1 & 0.985 & 0.009 & 0.006 & 1.020 & 0.008 & 0.941 & 0.001 & 0.007 & 1.019 & 0.001 \\
B2 & --    & --    & --    & --    & --    & 0.934 & 0.006 & 0.008 & 1.023 & 0.006 \\
B3 & 0.989 & 0.008 & 0.005 & 0.998 & 0.008 & 0.973 & 0.000 & 0.005 & 1.000 & 0.000 \\
B4 & --    & --    & --    & --    & --    & 0.971 & 0.005 & 0.005 & 1.005 & 0.005 \\
C1 & 0.979 & 0.004 & 0.008 & 1.024 & 0.003 & 0.920 & $-$0.010 & 0.009 & 0.997 & $-$0.010 \\
C2 & --    & --    & --    & --    & --    & 0.911 & $-$0.005 & 0.009 & 1.002 & $-$0.005 \\
C3 & 0.982 & 0.005 & 0.007 & 1.019 & 0.004 & 0.944 & $-$0.009 & 0.007 & 0.991 & $-$0.009 \\
C4 & --    & --    & --    & --    & --    & 0.935 & $-$0.003 & 0.008 & 0.996 & $-$0.004 \\
D1 & 0.983 & 0.007 & 0.007 & 1.020 & 0.007 & 0.943 & $-$0.007 & 0.007 & 1.005 & $-$0.007 \\
D2 & --    & --    & --    & --    & --    & 0.935 & $-$0.002 & 0.008 & 1.010 & $-$0.002 \\
D3 & 0.985 & 0.008 & 0.006 & 1.015 & 0.008 & 0.962 & $-$0.006 & 0.006 & 0.999 & $-$0.006 \\
D4 & --    & --    & --    & --    & --    & 0.956 & $-$0.001 & 0.006 & 1.004 & $-$0.001 \\
E1 & 0.953 & 0.013 & 0.012 & 1.038 & 0.012 & 0.916 & $-$0.021 & 0.009 & 1.022 & $-$0.022 \\
E2 & --    & --    & --    & --    & --    & 0.916 & $-$0.028 & 0.009 & 1.027 & $-$0.028 \\
E3 & 0.955 & 0.015 & 0.011 & 1.034 & 0.015 & 0.917 & $-$0.016 & 0.009 & 1.022 & $-$0.016 \\
E4 & --    & --    & --    & --    & --    & 0.916 & $-$0.022 & 0.009 & 1.028 & $-$0.023 \\
F1 & 0.960 & 0.011 & 0.011 & 1.023 & 0.010 & 0.960 & $-$0.036 & 0.006 & 0.998 & $-$0.036 \\
F2 & --    & --    & --    & --    & --    & 0.961 & $-$0.040 & 0.006 & 1.006 & $-$0.040 \\
F3 & 0.961 & 0.013 & 0.010 & 1.020 & 0.013 & 0.961 & $-$0.030 & 0.006 & 0.998 & $-$0.030 \\
F4 & --    & --    & --    & --    & --    & 0.963 & $-$0.034 & 0.006 & 1.006 & $-$0.035 \\
G1 & 0.993 & 0.012 & 0.005 & 1.030 & 0.011 & 0.991 & $-$0.001 & 0.003 & 1.016 & $-$0.001 \\
G2 & --    & --    & --    & --    & --    & 0.980 & $-$0.005 & 0.004 & 1.006 & $-$0.005 \\
G3 & 0.997 & 0.010 & 0.003 & 1.009 & 0.010 & 0.990 & $-$0.002 & 0.003 & 0.995 & $-$0.002 \\
G4 & --    & --    & --    & --    & --    & 0.995 & $-$0.007 & 0.002 & 0.987 & $-$0.007 \\
H1 & 0.976 & 0.020 & 0.008 & 1.044 & 0.019 & 0.929 & 0.010 & 0.008 & 1.030 & 0.010 \\
H2 & --    & --    & --    & --    & --    & 0.914 & 0.005 & 0.009 & 1.020 & 0.005 \\
H3 & 0.985 & 0.020 & 0.006 & 1.025 & 0.019 & 0.969 & 0.009 & 0.005 & 1.012 & 0.009 \\
H4 & --    & --    & --    & --    & --    & 0.961 & 0.004 & 0.006 & 1.003 & 0.004 \\
I1 & 0.967 & 0.013 & 0.010 & 1.044 & 0.012 & 0.764 & 0.007 & 0.018 & 1.079 & 0.007 \\
I2 & --    & --    & --    & --    & --    & 0.741 & 0.002 & 0.020 & 1.073 & 0.002 \\
I3 & 0.969 & 0.011 & 0.009 & 1.029 & 0.011 & 0.789 & 0.006 & 0.017 & 1.067 & 0.006 \\
I4 & --    & --    & --    & --    & --    & 0.767 & 0.002 & 0.018 & 1.062 & 0.002 \\
J1 & 0.960 & 0.023 & 0.011 & 1.063 & 0.022 & 0.752 & 0.017 & 0.019 & 1.093 & 0.016 \\
J2 & --    & --    & --    & --    & --    & 0.731 & 0.011 & 0.021 & 1.087 & 0.011 \\
J3 & 0.965 & 0.022 & 0.010 & 1.050 & 0.021 & 0.781 & 0.017 & 0.018 & 1.085 & 0.016 \\
J4 & --    & --    & --    & --    & --    & 0.760 & 0.011 & 0.019 & 1.079 & 0.011 \\
\hline
BT-Settl & 0.962 & 0.007 & 0.010 & 0.944 & 0.009 & 0.908 & $-$0.078 & 0.009 & 0.943 & $-$0.078 \\
GSL      & 0.917 & $-$0.034 & 0.015 & 0.976 & $-$0.033 & 0.805 & $-$0.045 & 0.013 & 0.898 & $-$0.045 \\
\hline
\end{tabularx}

\vspace{0.3em}
\begin{minipage}{\textwidth}
\footnotesize
Note. For $R^+_{\rm HK}$, linear comparisons were carried out only for the VALD-based configurations, i.e., labels 1 and 3. Entries marked with ``--'' therefore indicate combinations for which no $R^+_{\rm HK}$ linear fit was included.
\end{minipage}
\end{table*}

\FloatBarrier

\bibliography{references}{}
\bibliographystyle{aasjournal}
\end{CJK*}
\end{document}